\documentclass[10pt,letterpaper]{article}
\usepackage[top=0.85in,left=2.75in,footskip=0.75in]{geometry}

\usepackage{amsmath,amssymb}

\usepackage{changepage}

\usepackage{textcomp,marvosym}

\usepackage{cite}

\usepackage{nameref,hyperref}

\usepackage[right]{lineno}

\usepackage[nopatch=eqnum]{microtype}
\DisableLigatures[f]{encoding = *, family = * }

\usepackage[table]{xcolor}

\usepackage{array}

\usepackage{booktabs}
\usepackage{multirow}

\newcolumntype{+}{!{\vrule width 2pt}}

\newlength\savedwidth

\usepackage{placeins}

\raggedright
\usepackage[aboveskip=1pt,labelfont=bf,labelsep=period,justification=raggedright,singlelinecheck=off]{caption}

\makeatletter
\renewcommand{\@biblabel}[1]{\quad#1.}
\makeatother

\usepackage{lastpage,fancyhdr,graphicx}
\usepackage{epstopdf}
\fancyheadoffset[L]{2.25in}
\fancyfootoffset[L]{2.25in}
\begin{document}
\vspace*{0.2in}

\begin{flushleft}
{\Large
\textbf\newline{Dynamic signatures of collective knowledge through navigation in a network with semantic structure} 
}
\newline
\\
Abby Bertics \textsuperscript{1*},
Elizabeth R. Chrastil \textsuperscript{2},
Jean M. Carlson \textsuperscript{3}
\\
\bigskip
\textbf{1} Interdepartmental Graduate Program in Dynamical Neuroscience, University of California, Santa Barbara, CA 93106, USA
\\
\textbf{2} Department of Neurobiology \& Behavior, University of California, Irvine, CA 92697, USA
\\
\textbf{3} Department of Physics, University of California, Santa Barbara, CA 93106, USA
\\
\bigskip

%
%





* abertics@ucsb.edu

\end{flushleft}
\section*{Abstract} 
What distinguishes how humans navigate an information network, such as Wikipedia, from what the underlying graph structure and language statistics alone would predict? 
To address this question of human collective navigation, we analyze 70,770 human trajectories through Wikispeedia, an online game where players navigate from article to article to reach a target.
To compare groups of trajectories rather than individual ones, we merge all trajectories leaving or entering a given article into a single data structure: a Flow Tree. Flow Trees compress collections of trajectories into hierarchies of group splits. We build forward and reverse Flow Trees for 815 of the most commonly traversed articles, representing divergence and convergence, respectively. 
We compare humans against four random-walker null models, whose decisions are either unweighted or weighted by graph structure, word frequency, or semantic similarity to target. 
Human Flow Trees are deeper with longer trunks and later splits, indicating players stay together longer than random walkers, whose Flow Trees tend to fan out immediately. 
An article's degree predicts 51.3-70.9\% of the variation in random-walker Flow Tree branching factor, but only 24.1-36.2\% of human Flow Tree branching factor. 
Taking what graph structure fails to predict as a two-dimensional coordinate system, we define four article types (beacons, funnels, forks, and corridors), distinguished by how humans enter and exit them. 
These human-defined types are not explained by out-degree or by how common the title words are, and are not recovered by any random walker. 
A fifth null model, built from first-order human click preferences, partially reproduces human Flow Tree shape but differs significantly from humans in how much of that shape degree explains.
We consider \textit{collective knowledge} to be the structure shared by players who never communicate, beyond what graph structure imposes. It comprises first-order click preferences and a significant component that extends across clicks.

\clearpage
\newgeometry{top=0.85in,left=1in,right=1in,footskip=0.75in}

\section*{Introduction}
In public discourse, ``the wisdom of the crowds'' refers to the idea that groups of individuals tend to make better decisions than even the smartest of the bunch~\cite{galton1907one,surowiecki2004}. Aggregation does more than improve accuracy. Sometimes individuals do the same thing, even though they have never met and no one has told them what to do or how to do it. Much to the chagrin of humans, ants collectively converge on an efficient route to stray crumbs~\cite{vittori2006path}. Lawns on college campuses across the world have diagonal dirt paths across their corners, reflecting how short-cutters converge on similar routes~\cite{helbing2001self}. Strangers, asked to give directions in a city, might use similar landmarks, or anchors, to explain the way~\cite{lynch1960image,couclelis1987exploring}. In all of these cases, there is something that exists in the collection that is not coordinated by nor does it exist within a single individual.

Navigation renders this explicit and sometimes, like in the case of dirt paths, visual. When many people navigate through the same environment independently, their aggregated choices may encode information beyond that of an individual. We will refer to structure of this kind, shared across individuals and exceeding that imposed by the environment, as \textit{collective knowledge}. Below, we operationalize this as structure in human navigation that is beyond what our models of graph structure and language statistics predict. Similar notions have appeared across fields under various names and guises.
In psychology, there is the concept of ``collective intelligence'', through which group performance on cognitive tasks is correlated with social sensitivity, not individual IQ~\cite{woolley2010evidence}; aggregation that exploits individual heterogeneity is key for it to emerge~\cite{kameda2022information}. 
In Wikipedia navigation specifically, the differences in strategies of many players attempting the same game shed light on the information landscape around the target~\cite{zhu2025milgram}.
In cognitive anthropology, distributed cognition is the idea that thinking and decision-making are done by a system of people and instruments rather than by any single entity. This was first analyzed in the context of naval navigation, where no one sailor can maintain the ship's position, but rather it is maintained through the coordinated activity of the crew and their instruments~\cite{hutchins1995cognition}.
Similar ideas appear in human-computer interaction~\cite{hutchins2000distributed,perry2003distributed}, transactive memory system theory~\cite{wegner1987transactive}, organization theory~\cite{lam2000tacit}, and educational learning theory~\cite{bereiter2002education,cress2018collective}.

Hecker sought to taxonomize all the definitions of collective knowledge used in organization theory, and settled on three: the intersection of individual knowledge, the union of individual knowledge, and the knowledge that is embedded in external artifacts, such as technology, rules, or systems~\cite{hecker2012knowledge}. Each of these notions differs in what it takes ``collective'' to mean, in how it defines ``knowledge'', and in the mechanism, if any, by which one produces the other. We do not seek an authoritative definition in this work, but rather we seek to carve out a slice that we can measure quantitatively.

We do this by considering navigation. Most of the time collective knowledge is inferred from the end result that a group produces, be that a decision or solution. Navigation is in essence a sequence of decisions, as such capturing the process, in addition to the final result. When many people navigate the same environment independently, their aggregated choices are observable as they progress from start to end. Moreover, this navigation need not be spatial. There is well-established evidence showing that humans move through conceptual structures using very similar cognitive machinery to that which they use in spatial navigation, with knowledge organized in graph-like relational maps that can be searched and traversed~\cite{chrastil2014cognitive,peer2021structuring}. 
 
Knowledge graphs on the internet, such as Wikipedia, make such a space concrete. They have both graph structure (the set of article vertices and edge hyperlinks connecting them) that constrains which directions navigators can click to go next and semantic structure (conceptual relationships between articles) that together influence navigation decisions and strategy~\cite{west2009wikispeedia,livingstone2016models}. Wikipedia itself was collectively constructed, and this structural–semantic duality has been used to extract collective knowledge from how people move across it, inferring which concepts are close together and which are far~\cite{singer2013computing}.
In the game of Wikispeedia~\cite{west2009wikispeedia}, players must navigate from a starting article to a target article by clicking only on links within Wikispeedia (a subset of the larger Wikipedia graph). 
Wikispeedia is a research implementation of a widely played game, which has gone by various names, such as the Wikipedia Game, the Wiki Game, or wikiracing~\cite{thewikigame}. Navigation data from one such public site has also been used to study memory in human navigation~\cite{singer2014detecting}. More recently, the game has become a benchmark for automated navigators, from reinforcement-learning agents~\cite{nogueira2016end} to sentence-embedding models and large language models (LLMs)~\cite{west2012automatic,margiotta2025evaluating,ebrahimi2025textual,ziomek2026llm,hickey2026can}.
Prior work analyzing this dataset has shown that human players can efficiently navigate the graph, but often fail to use the shortest paths, instead following semantically meaningful routes between concepts~\cite{west2012human}. West and Leskovec conjectured the use of certain articles to be instrumental in the games of human players~\cite{west2012human}. Players first ``zoom out'' to these key articles before ``homing in'' on the target. 

We investigate how players diverge from and converge on articles using Flow Trees, a new tool introduced in~\cite{bertics2026flowtree}. Take every player who has passed through a given article and follow them forward. Many might click to the same next article, some even staying together through multiple clicks. A Flow Tree records how trajectories diverge; a long trunk means players stay together and are more cohesive, while immediate fan-like branching means players go their separate ways early. Built in reverse, this same structure represents how players converge to an article. 
The Flow Tree is related to the Reeb graph from computational topology~\cite{reeb1964formes}, which in a similar manner summarizes how level sets of a function merge and split in topological structure as they evolve over a manifold.
Flow Trees compress a static hyperlink structure into a representation of dynamic flow, which lets us ask not only which articles are structurally central, but also which ones disproportionately shape where players go. The two can differ, because article use depends on context. The same article may be used one way in some games and another in others, depending on the goal. This dependency is not visible in static link structure.

The difficulty in interpreting any such measure is that the underlying Wikispeedia graph alone produces structure in navigation. An article with many outgoing links will scatter any navigator, human or otherwise. We compare human Flow Trees against four random-walker null models that traverse the same graph from the same starting articles, differing only in how each decision point is weighted: uniformly, by neighbor degree, by neighbor word frequency, or by semantic similarity to the target. Comparing human navigation against synthetic navigators has precedent, including decentralized-search models~\cite{helic2013models} and automatic agents on Wikispeedia itself~\cite{west2012automatic}.
The null models represent navigation driven by graph structure or by language statistics alone. Anything humans do that is not reproduced by the random walkers can therefore not be attributed to these models of graph structure and language statistics. A fifth null model, built from human click frequencies but with no memory and no target, asks how much of what remains is determined by single clicks. 
Walks of this kind have been used to approximate human Wikipedia navigation~\cite{arora2022wikipedia}; we use this as a baseline, so that whatever structure it fails to reproduce must extend across clicks.

Our contributions are: 
(1) we apply Flow Trees to Wikispeedia trajectories to quantify how much of human navigation data is explained by graph structure and how much is not; 
(2) we introduce a two-dimensional description of an article's navigational role, and use it to define four article types; 
(3) we show that these roles are partially reproducible across independent halves of the human data and are not recovered by any random walker;
and (4) using a history-free model of human clicks, we show that human departure from graph structure is partially explained by first-order click preferences, but that a significant component extends across clicks.

\section*{Materials and Methods}
\subsection*{Wikispeedia dataset}
\begin{figure}[!ht]
\centering
\includegraphics[width=\textwidth, page=1]{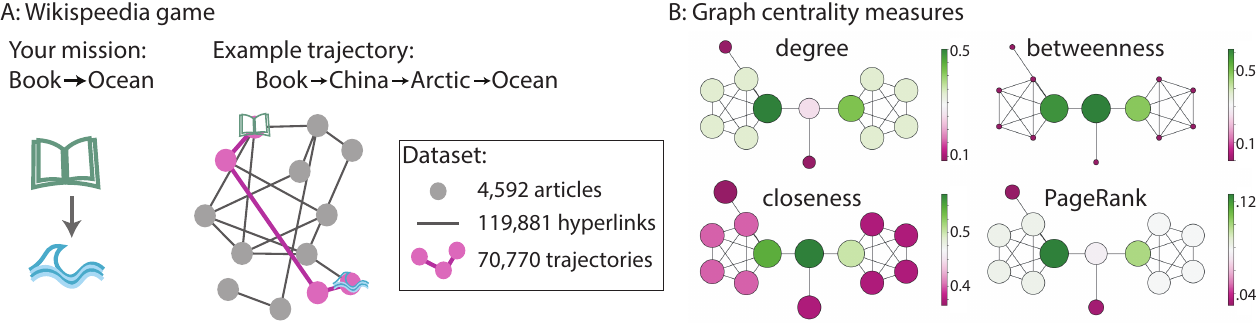}
\caption{\textbf{Wikispeedia dataset and graph.}
A: In the game of Wikispeedia~\cite{west2009wikispeedia} players navigate from a randomly assigned start article to a target article by following hyperlinks within a curated subset of Wikipedia (4,592 articles, 119,881 directed edges).
We show an example game (start: ``Book'', target: ``Ocean'') and a corresponding example trajectory (``Book''$\to$``China''$\to$``Arctic''$\to$``Ocean'') through a cartoon depiction of the Wikispeedia graph, where articles are nodes and hyperlinks from one article to another are edges.
B: The graph nature of Wikispeedia allows us to use common graph measures (degree, betweenness, closeness, and PageRank centrality) to quantify how articles might be impacted by graph structure. We show a toy example, with nodes colored and sized so that big and dark green are the most central and small and dark pink are the least. Different measures provide different orderings of nodes.}
\label{fig:wikispeedia}
\end{figure}

To test ideas of collective knowledge in graph-like structures, we use the Wikispeedia dataset collected in~\cite{west2009wikispeedia}. The original purpose of this dataset was to use crowd-sourced Wikipedia navigation data to compute the semantic distances between pairs of concepts. In this game, players must navigate from one article to another, as fast as they can, only clicking on links in the articles (see Fig~\ref{fig:wikispeedia}A).  

The underlying Wikispeedia graph (which we will interchangeably refer to just as Wikispeedia) is a subset of Wikipedia, with 4,592 articles (nodes; we will use article for Wikipedia pages and node for Flow Tree nodes) and 119,881 hyperlinks (edges) between them. The dataset is composed of player trajectories from 76,193 games, which start at 4,168 and end at 3,998 unique articles. We include both finished and unfinished games~\cite{scaria2014last}. To limit repeat attempts, we keep each player's first game from each start article. This leaves us with a total of 70,770 trajectories for our analyses below.

\subsubsection*{Centrality measures}
Wikispeedia's dual representation as a graph and as text allows us to consider both the structural centrality (see Fig~\ref{fig:wikispeedia}B) and the semantic centrality of articles. For structural centrality, we use the following measures: \textit{degree} centrality (the number of edges incident to each article; we use
in-degree and out-degree separately when direction matters), \textit{betweenness} centrality (shortest paths passing through each article), \textit{closeness} centrality (inverse of the mean shortest-path distance to all other articles), and \textit{PageRank} centrality (stationary probability of a random walk on the graph~\cite{brin1998anatomy}). We use the \texttt{networkx} Python package~\cite{hagberg2008exploring}.

For semantic centrality, we use the mean log-scaled frequency\footnote{Word frequencies are approximately power-law distributed, spanning several orders of magnitude, known as Zipf's Law~\cite{zipf1945meaning}, so a raw-frequency measure would be dominated by a handful of semantically vacuous, function-like words; we therefore work on the log scale~\cite{van2014subtlex}, where psycholinguistic quantities such as lexical decision time vary approximately linearly~\cite{brysbaert2009moving}.} of the words in an article's title as a proxy. More frequent words appear across a wider range of contexts than rarer ones \cite{adelman2006contextual}, so the concepts they represent are both more widely known and easier to relate to other topics and articles, which makes them reasonable navigational tools for a navigator without a priori knowledge of the global underlying Wikispeedia graph structure. 

\subsection*{Flow Trees}
\begin{figure}[!ht]
\centering
\includegraphics[width=\textwidth, page=2]{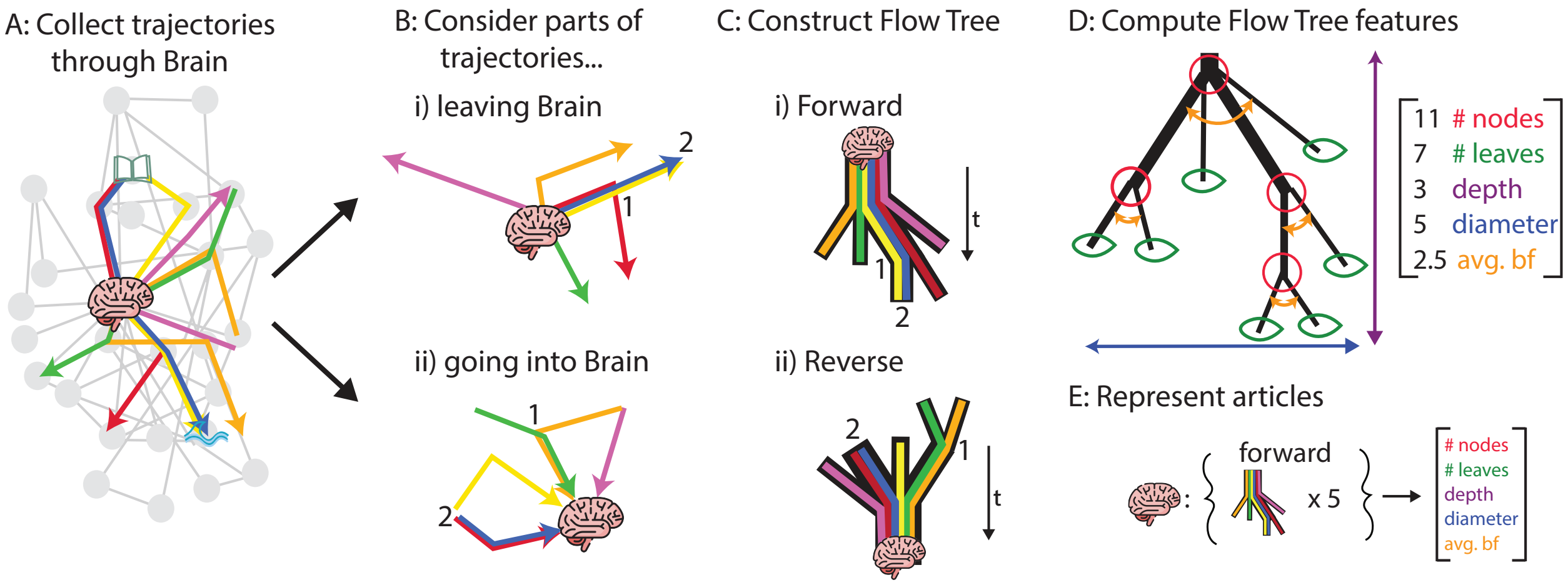}
\caption{\textbf{Flow Tree construction in Wikispeedia.}
A: We first collect all trajectories that pass through a given article, e.g., ``Brain''.
B: We then consider segments of these trajectories. i) For flow out, we collect all trajectories with at least five decisions after leaving the article and truncate them to length five. ii) Same for flow in, but prior to arriving at the article. 
C: Flow Tree construction~\cite{bertics2026flowtree}. i) \textit{Forward} Flow Trees encode the structure of outgoing trajectories. Flow Tree nodes correspond to places where the subgroups of trajectories diverge from each other on the navigation graph (see 1 and 2).
ii) \textit{Reverse} Flow Trees are built from time-reversed trajectories and encode the structure of incoming flow. 
D: We quantify each Flow Tree using standard tree features: number of nodes, number of leaves, depth, diameter, and average branching factor. See~\nameref{sec:ft-feats} and \cite{bertics2026flowtree} for more detail.
E: We represent the divergence dynamics from each of the articles as a five-dimensional vector, which is the average forward Flow Tree features over five sampled trees. The same procedure applies for reverse Flow Trees. 
}
\label{fig:howtobuild}
\end{figure}

Flow Trees summarize the collective dynamics of a group of trajectories by encoding when groups of trajectories diverge (see Fig~\ref{fig:howtobuild}A-C). They represent the hierarchical splitting structure of trajectories through a space. Details about construction, as well as analysis for a human maze navigation dataset were introduced in \cite{bertics2026flowtree}. In that study, we built \textit{path} Flow Trees from groups of trajectories that share both a starting point and a target, using trajectories of humans navigating a virtual-reality maze, and showed that more difficult navigation tasks yield more scattered Flow Trees. The path Flow Trees used in that study require many navigators to solve the same problem, which is possible in a laboratory design, but was not the case with Wikispeedia. Its games were spread across around 40,000 start-target pairs, with a median of one game per pair; only 96 pairs were played at least 15 times. In this work, we construct \textit{article} Flow Trees, which consider all trajectories entering or exiting a given article, regardless of its start or target. Here, we briefly recapitulate the Flow Tree framework. 

\subsubsection*{Definition}
A Flow Tree $FT=(V_{FT},E_{FT},r,a)$ is a rooted, directed tree whose topology reflects the order in which trajectories diverge. More formally, it is a connected acyclic graph with a root node $r\in V_{FT}$ such that every node is reachable from $r$ by a unique directed path. Edges point away from the root. The attribute map, $a: E_{FT}\to (0,1]$ assigns a weight to each edge, representing the proportion of trajectories following that branch.

\subsubsection*{Construction}

Flow Trees are constructed from a set of trajectories $\mathcal{T}=\{T^i\}$, where each trajectory
$T^i=(t^i_0,t^i_1,\dots,t^i_{n_i-1})$ is a sequence of decisions or, equivalently, visited articles (Fig~\ref{fig:howtobuild}A-C). All trajectories start together at the root. Construction proceeds recursively. At each step, trajectories are grouped according to their next decision. If the same next article is chosen unanimously, the tree does not branch. But if multiple distinct decisions occur, the group splits and a node is created for each subgroup. The procedure continues recursively for each subgroup until trajectories finish or each subgroup contains a single trajectory. 

\subsubsection*{Features}
To quantify and compare Flow Trees, we represent each Flow Tree as a five-dimensional vector of standard tree metrics (number of nodes, number of leaves, depth, diameter, and average branching factor; see Fig~\ref{fig:howtobuild}D and~\nameref{sec:ft-feats} for more detail)~\cite{bertics2026flowtree}. Since there are often more trajectories than the amount required to construct one Flow Tree, we sample the requisite amount five times to construct five Flow Trees per article, compute the feature vector for each tree, and then average across the Flow Tree samples per article, resulting in article-level means (see Fig~\ref{fig:howtobuild}E).

\subsubsection*{Hyperparameters}

The hyperparameters relevant for Flow Tree construction are related to the groups of trajectories, and the optimal choice depends on the sparsity of graph and availability of data. For the Wikispeedia dataset, we collect groups of 5-decision-long trajectories starting from each article. Longer trajectories allow for more divergence, but requiring longer trajectories disqualifies some data as it requires at least five decisions before or after the article in question. After trying $n\in\{3,4,5,6\}$, we settled on five to permit maximum structure, while still allowing for an adequate number of trajectories to sample from. Then, for each of these groups, we sample 15 trajectories five times to construct Flow Trees. 

We consider the 1000 most commonly traversed articles and only keep those with at least 15 incoming trajectories of length 5 and at least 15 outgoing trajectories of length 5. This leaves us with 815 articles for downstream analysis.

\subsection*{Null models}
\subsubsection*{Unbiased, graph-structure, semantic, and targeted random walkers}
\label{sec:matmeth-rw}
\begin{figure}[h!t]
\centering
\includegraphics[width=\textwidth, page=3]{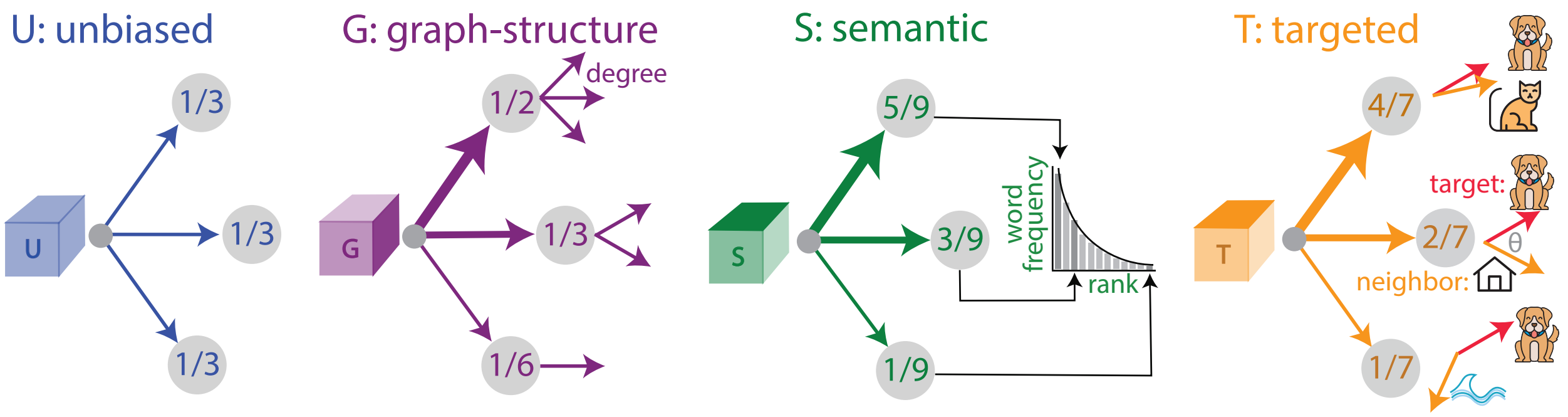}
\caption{\textbf{Random-walker null models.} At a given article (small gray circle) a walker samples its next article from its outgoing links, weighted according to one of four rules. Weights are normalized to sum to one to act as a probability distribution. \textit{Unbiased (U)} random walkers distribute probability equally across all outgoing links. \textit{Graph-structure (G)} random walkers weight each link by the out-degree of the neighboring article, so that links to well-connected pages are preferred. \textit{Semantic (S)} random walkers weight each link by the average word frequency (in English) of the words in the article's title. This biases the walker towards articles with titles with more common words (e.g., ``Science'', rather than ``Differential Geometry''), a proxy for more general concepts. \textit{Targeted (T)} random walkers weight links based on how semantically similar the neighbor's title is to the target's title. For instance, ``Dog'' is closer to ``Cat'' than it is to ``House'' or ``Ocean''. ($\theta$ is the angle between embedding vectors.)}
\label{fig:randomwalkers}
\end{figure}

In order to isolate human collective knowledge from the effects of graph and language structure, we develop four different null models that we compare to humans. Each null model is a random walker that creates a set of trajectories, matched to the start articles and trajectory lengths of the human trajectories. All four random walkers select neighbor $n_i$ with probability $P(n_i)$. What changes is the weighting of the $N$-sided die (see Fig~\ref{fig:randomwalkers}). 

\begin{itemize}
    \item \textbf{Unbiased random walker (U)}: selects the next article from its $N$ neighbors uniformly at random (i.e., using an unbiased die),
    $$P(n_i)=\frac{1}{N}$$
    \item \textbf{Graph-structure random walker (G)}: uses a die weighted according to each neighbor's out-degree on the graph being traversed, 
    $$P(n_i) = \frac{\text{out-degree}(n_i)}{\sum_j \text{out-degree}(n_j)}.$$
    \item \textbf{Semantic random walker (S)}: uses a die weighted according to each neighbor's mean word frequency of the words in its title, excluding stop words (a, an, the, and, for, and of). We use the \texttt{wordfreq} Python library~\cite{robyn_speer_2022_7199437}. This walker preferentially visits articles with more common words in its title, independent of the target,
    $$P(n_i) = \frac{z(n_i)}{\sum_j z(n_j)},$$
    where $z(x)$ is the average Zipf frequency (base-10 logarithm of the number of times a word appears per billion words) of the words in $x$'s title~\cite{van2014subtlex}.
    \item \textbf{Targeted random walker (T)}: uses a die weighted according to the cosine similarity between the title embedding~\cite{devlin2018bert} of each neighbor and that of the target article. At page $p$ with neighbors $n_1,\dots,n_k$ and a fixed target $t$, we do the following:
    \begin{enumerate}
        \item Tokenize and embed each title in $\{n_1,\dots,n_k\}
        \cup \{t\}$ with \texttt{bert-base-uncased} and \texttt{AutoTokenizer} from the Hugging Face \texttt{transformers} Python library \cite{wolf-etal-2020-transformers}. This results in a vector representing the semantics of each article: $\vec{v}_{n_1},\dots,\vec{v}_{n_k}$ and $\vec{v}_t$.
        \item Compute the cosine similarity between the target and each neighbor's title embedding vectors:
        $$\mathrm{s}(n_i,t) = \cos\theta_i = \frac{\vec{v}_{n_i}\cdot\vec{v}_t}{\|\vec{v}_{n_i}\|\,\|\vec{v}_t\|}.$$
        \item Normalize into a probability distribution over neighbors: $$P(n_i) = \frac{\mathrm{s}(n_i,t)}{\sum_j \mathrm{s}(n_j,t)}.$$
    \end{enumerate}
    Targeted semantic random walkers are matched to humans on target as well as start and length.
\end{itemize}

The four random walkers above differ in how strongly they concentrate decisions at each click. We measure this as the normalized entropy of the distribution, $H/\log{N}$ where $H = -\sum_i P(n_i)\log P(n_i)$. The unbiased random walker has a normalized entropy of one, since every link is equally likely. Averaged over articles, the semantic and targeted random walkers are nearly uniform ($H_S=0.994$ and $H_T=0.996$), and the graph-structure random walker is more selective ($H_G=0.927$).

\subsubsection*{Reverse trajectories and reverse Flow Trees}
We construct human reverse Flow Trees by reversing the human trajectories and applying the forward Flow Tree construction process. We construct random-walker reverse Flow Trees in a parallel manner: we first generate forward walks on the Wikispeedia graph, matched in length, start, (and, for the targeted random walker, target). Then, we reverse these trajectories and construct Flow Trees. Another option would have been to skip the forward trajectory generation step and directly generate reverse trajectories by using the random walkers on the reversed Wikispeedia graph. The two are not equivalent (see~\nameref{sec:forward-reverse-dynamics}), and we chose the former, since it is the same as how we treat the human data.

\subsubsection*{History-free human model}
\label{sec:matmeth-hf}
We introduce a fifth null model to determine whether human collective knowledge reduces to first-order click preferences, or whether it contains extended temporal structure. 
Whether human web navigation is memoryless is a long-standing question. Higher-order Markov models are favored based on likelihoods~\cite{chierichetti2012web}, but first-order models appear sufficient at the level of individual pages given available data, and memory appears only when pages are grouped into topics~\cite{singer2014detecting}. Whether there is memory has an impact on downstream analyses, such as inferring community structure~\cite{rosvall2014memory}. We do not fit Markov models of increasing order, but rather ask whether a first-order model reproduces the shape of collective human navigation.

The history-free human model is a Markov model of human clicks. It reduces the human Wikispeedia navigation data to a first-order transition matrix with no memory of previously visited articles. For instance, suppose ``Ice cream'' only links to ``Chocolate'', ``Strawberry'', and ``Vanilla'', and in the human data they have been clicked on 12, 8, and 4 times, respectively. Then, our history-free human model will have a $\frac{1}{2}$ chance of going to ``Chocolate'', a $\frac{1}{3}$ chance of going to ``Strawberry'', and a $\frac{1}{6}$ chance of going to ``Vanilla''. To do this in practice, we split players hashed by IP address into two halves, and on the first half, we use maximum likelihood estimation with additive smoothing ($\alpha=0.1$ 
fake clicks added to every outgoing link so that links never clicked on get a small probability) to compute the transition probability matrix; from current article $c$, this makes the probability distribution over its neighbors: 
$$P(n_i|c)=\frac{\text{\#clicks from }c\text{ to }n_i + \alpha}{\text{\#clicks from }c + N\alpha},$$
where $N$ is the number of outgoing links of $c$.

Arora et al.\ similarly generated synthetic Wikipedia reading sequences from aggregate click counts, by using first-order random walks whose transition probabilities are proportional to how often each link was clicked, matched to real sequences in starting article and length~\cite{arora2022wikipedia}. They asked whether such walks are an adequate substitute for real navigation data. We instead use these walks as a null model to determine the existence of temporally extended structure in human Wikispeedia navigation. Our implementation differs in three ways: we estimate transition probabilities from goal-directed Wikispeedia games, rather than free browsing; we smooth transition probabilities so all links maintain a small nonzero probability; and we estimated them from one half of players and compare them against the other half.

This history-free human model represents the first-order click preferences of humans, but has no memory of previously visited articles and no target. This is not equivalent to goal-free, though, as its transition probabilities are estimated from players who each have a goal. It will therefore inherit whatever goal-directedness is present in aggregate single-click statistics. All comparisons with this model will use only the second, held-out half of the human data, with random walks matched to the starts and lengths of these held-out trajectories. Since these results depend on the random split, we repeat the following analysis over five random seeds and report means with ranges. Requiring enough held-out trajectories restricts these comparisons to 488-539 articles for forward analysis and 611-663 for reverse for which there is sufficient trajectory data, depending on the random split.

\subsection*{Regression and residual analysis}
To determine how much of an article's Flow Tree shape is explained by its connectivity, we regress each forward Flow Tree feature against log out-degree and each reverse Flow Tree feature against log in-degree, using SciPy's \texttt{linregress}. We also fit multivariate models including both in- and out-degree to confirm that the non-dominant direction has negligible contribution.

The residual for each article is the observed feature value minus the regression prediction, and measures how far that article's Flow Tree departs from what its degree alone predicts. By construction each residual is orthogonal to its own predictor. We use the forward and reverse branching-factor residuals as a pair of coordinates and classify articles by the quartiles of each axis. Articles in the upper quartile of both, the lower quartile of both, or opposite quartiles, define four corner types. To verify that residuals reflect reproducible article properties rather than sampling noise, we split each article's trajectories into disjoint halves, build Flow Trees from each half, compute residuals independently, and correlate them across articles. This requires at least 30 trajectories, enough to make two Flow Trees, in each direction. To fairly compare humans to random walkers, we first subsample each navigator's trajectories at each article to match the smallest amount among the five navigators. We repeat this over five random seeds.

To compare how consistent the history-free human model is with humans in terms of residuals, we also build Flow Trees from the trajectories of the training players and correlate their degree residuals with those of the held-out players. This human-human correlation acts as a ceiling and is possible for 451–482 articles (forward) and 582–612 (reverse).

\subsubsection*{Statistical analysis}

We compare correlations that share a variable (e.g., human traversal counts correlated with a centrality measure and random-walker traversal counts correlated with that same measure) with Steiger's $Z$-test for dependent correlations. We compare Flow Tree features between humans and each walker across articles using Welch's $t$-test, with Cohen's $d$ as effect size. We correct all $p$-values within a table using the Benjamini–Hochberg procedure. Confidence intervals on Pearson correlations use the Fisher $z$-transformation; those on differences in $R^2$ use 10,000 bootstrap resamples of articles. Agreement between type assignments is measured with Cohen's $\kappa$. Analyses that depend on a random split or subsample are repeated over five random seeds. Brackets denote a range over five seeds unless it is labeled as a $95\%$ confidence interval.

\subsection*{Data accessibility}
The Wikispeedia dataset is available at \url{https://snap.stanford.edu/data/wikispeedia.html}. We accessed it for research purposes on 29 October 2025. Players are identified only by hashed IP address, and the authors had no access to information that could identify individual participants during or after data collection. Code to reproduce analysis and figures is available at \url{https://github.com/geometric-intelligence/mazedyn}.
 
\section*{Results}
We report six analyses, each of which builds upon the one before. In~(1) we ask whether human traversal of the Wikispeedia graph can be explained by graph or semantic structure alone, by correlating article traversal counts against semantic and structural centrality measures for humans and the four random-walker null models. 
In~(2) we replace traversal counts with Flow Trees, which retain information about the dynamics of collective navigation, and compare the divergence and convergence structure of human and random-walker trajectories. We use \textit{divergence} to mean how a group of trajectories leaving a given article splits apart over the following clicks, and \textit{convergence} to mean how a group arriving at that article comes together over the preceding clicks.
In~(3) we ask how much of Flow Tree structure is explained by out-degree and in-degree. 
In~(4) we use what degree does not explain as a two-dimensional coordinate system to describe how an article is used relative to how it is connected. Partitioning this space gives us four article types. 
In~(5) we test whether those types are reproducible and whether any random walker recovers them. 
In~(6) we evaluate the extent to which structure in human trajectories is temporally extended in nature.

\subsection*{(1) Human traversal counts track structural centrality less than any random walker, and semantic centrality more than all but the semantic random walker}
Both semantic and structural cues could influence human navigation decisions in the Wikispeedia game. Indeed, the two are correlated in the Wikispeedia graph; general concepts are also usually well connected. This correlation means the two influences cannot be separated by only considering human behavior. We use the null models as leverage to pry them apart. Each random walker is driven by exactly one of these cues and traverses the same graph from the same starting articles. Comparing humans against these simplified counterpoints allows us to consider how human behavior compares for each influence, even as they remain correlated in the underlying Wikispeedia graph.

\begin{figure}[!ht]
\centering
\includegraphics[width=\linewidth, page=4]{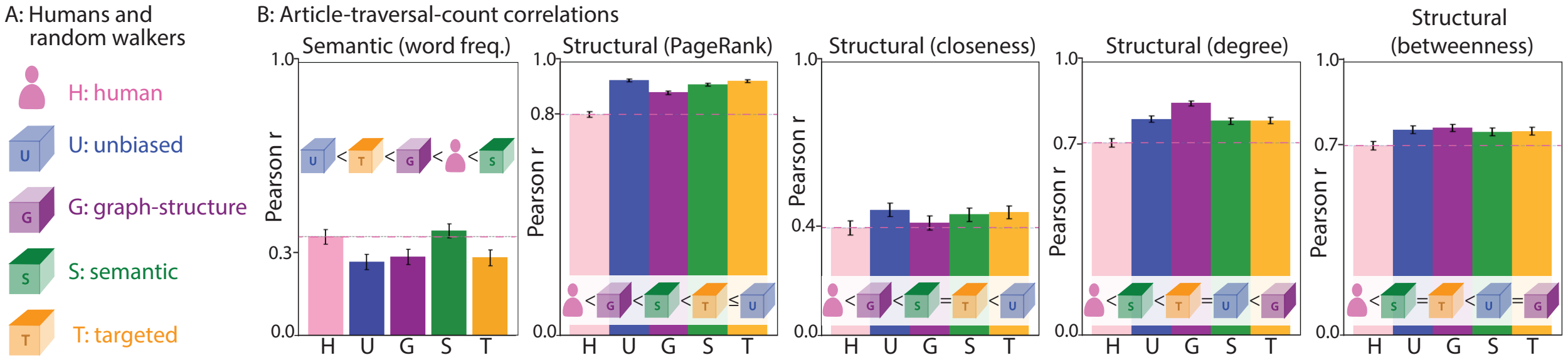}
\caption{\textbf{Semantic and structural influences on human navigation.}
A: Four null models used to disentangle human collective knowledge from graph and semantic structure: unbiased (U, blue) and three biased by graph centrality (G, purple), semantic centrality (S, green), and targeted semantic distance (T, orange). See Fig~\ref{fig:randomwalkers} and~\nameref{sec:matmeth-rw} in Materials and Methods for more details.
B: Pearson correlations of traversal counts with semantic centrality (mean log-scaled frequency of article title words, left) and structural centrality measures (PageRank, closeness, degree, and betweenness centrality, right) for humans and each random walker type. Error bars show 95\% confidence intervals on each Pearson correlation, computed via the Fisher z-transformation. Inequalities are embedded in the histograms to illustrate ordering. For the semantic centrality measure, the ordering of correlations is $S > H > \{G, T, U\}$; humans are more semantically driven than the random walkers that ignore word frequency, and less semantically driven than the random walker that is biased directly by it. For every structural measure, human traversal counts are less correlated than all four random walkers (see Table~\ref{tab:correlations} for statistics).
}
\label{fig:humanvsrandomstatic}
\end{figure}

The first analysis attempts to isolate these effects of structural and semantic centrality. For every article we record the number of times it is traversed by humans and by each of four random-walker null models (Fig~\ref{fig:humanvsrandomstatic}A). We then correlate these five sets of traversal counts against the one semantic and four structural measures of article centrality discussed in Materials and Methods, and use Steiger's $Z$-test to determine the extent to which the human correlation differs from that of each random walker.

\begin{table}[h!]
\centering
\caption{Pearson correlations of traversal counts with semantic and structural centrality measures for humans and four random-walker models.
Steiger's $Z$-test tests whether human traversal counts are more or less correlated with each measure than each random walker; a positive $Z$ indicates humans are more correlated. 
The highest correlation in each row is in \textbf{bold}. 
H=human, U=unbiased random walker, G=graph-structure random walker, S=semantic random walker, T=targeted random walker. 
$^{***}$ $p<0.001$, $^{**}$ $p<0.01$; $^{*}$ $p<0.05$; $p$-values are Benjamini–Hochberg corrected across the 20 comparisons.}
\small
\begin{tabular}{lccccccccc}
\toprule
 & \multicolumn{5}{c}{Pearson $r$} & \multicolumn{4}{c}{Steiger $Z$ (H vs.)} \\
\cmidrule(lr){2-6}\cmidrule(lr){7-10}
Centrality Measure & H & U & G & S & T & U & G & S & T\\
\midrule
\textit{Semantic} \\
\hspace{0.2cm}Log word frequency
  & 0.360 & 0.267 & 0.286 & \textbf{0.382} & 0.283
  & $+14.2^{***}$ & $+9.6^{***}$ & $-3.5^{***}$ & $+12.1^{***}$ \\
\midrule
\textit{Structural} \\
\hspace{0.2cm}Betweenness 
  & 0.693 & 0.757 & \textbf{0.761} & 0.747 & 0.751
  & $-12.8^{***}$ & $-12.1^{***}$ & $-11.4^{***}$ & $-11.9^{***}$ \\
\hspace{0.2cm}Closeness 
  & 0.391 & \textbf{0.462} & 0.415 & 0.442 & 0.452
  & $-10.5^{***}$ & $-3.0^{**}$ & $-7.9^{***}$ & $-9.2^{***}$ \\
\hspace{0.2cm}Degree 
  & 0.705 & 0.793 & \textbf{0.852} & 0.785 & 0.790
  & $-20.8^{***}$ & $-34.7^{***}$ & $-19.6^{***}$ & $-20.6^{***}$ \\
\hspace{0.2cm}PageRank 
  & 0.807 & \textbf{0.936} & 0.891 & 0.920 & 0.933
  & $-54.4^{***}$ & $-22.8^{***}$ & $-44.2^{***}$ & $-53.0^{***}$ \\
\bottomrule
\end{tabular}
\label{tab:correlations}
\end{table}

Human traversal counts are more strongly correlated with semantic centrality than are the counts of the unbiased ($Z=14.2$, $p<0.001$), graph-structure ($Z=9.6$, $p<0.001$) and targeted ($Z=12.1$, $p<0.001$) random walkers, and less strongly correlated than those of the semantic random walker ($Z=-3.5$, $p<0.001$; see Fig~\ref{fig:humanvsrandomstatic}B and Table~\ref{tab:correlations}). This ordering of correlation strengths, where humans are in the middle between graph-structure and semantic random walkers, is what would be expected if word frequency is one of several signals that humans use. To one extreme, the semantic random walker explicitly uses the semantic centrality measure to bias its choices, and therefore sets an upper bound on how correlated a navigator's counts can be with it. To the other, the unbiased random walker ignores semantics altogether (beyond what is implicit in Wikispeedia's hyperlink structure) and thus sets a lower bound. The unbiased random walker's correlation ($r=0.267$) reflects the residual signal that arises due to the confound that articles with frequently used title words tend to also be well-connected. Humans lie between these two bounds, indicating that they are sensitive to title word frequency as a measure of semantic centrality, but that their decisions are not only driven by that.

The graph-structure random walker uses no semantic information and reaches $r=0.286$, slightly above the unbiased random walker, confirming that degree in the Wikispeedia graph is a useful indicator of a concept's semantic centrality. The targeted random walker shows something slightly more nuanced. It is semantically biased, but towards articles close to the target article in abstract embedding space, rather than naively just towards frequent words. Its correlation ($r=0.283$) is indistinguishable from that of the unbiased random walker. This suggests a navigator can be semantically driven and still produce no word-frequency signature. Therefore finding that humans are more correlated than three of the random walkers suggests that humans use concept familiarity to at least partially guide their decision-making.

The structural centrality measures show humans in no such intermediate position. Human traversal counts are less strongly correlated with betweenness, closeness, degree, and PageRank centrality than are the counts of every
random walker. All sixteen Steiger comparisons are negative and significant (Fig~\ref{fig:humanvsrandomstatic}B, Table~\ref{tab:correlations}). The smallest margin is against the graph-structure random walker on closeness centrality ($Z=-3.0$, $p=0.003$); the largest is against the unbiased random walker on PageRank centrality ($Z=-54.4$, $p<0.001$). The unbiased random walker's high correlation with PageRank is expected, since its traversal counts are a finite-sample estimate of the stationary distribution that PageRank itself also approximates. Humans are therefore not simply following clues in the underlying Wikispeedia graph's structure. Subsequent analyses investigate what humans are doing instead.
 
\subsection*{(2) Human Flow Trees are deeper and less fan-like than those of any random walker}
\begin{figure*}[!ht]
\centering
\includegraphics[width=\linewidth, page=5]{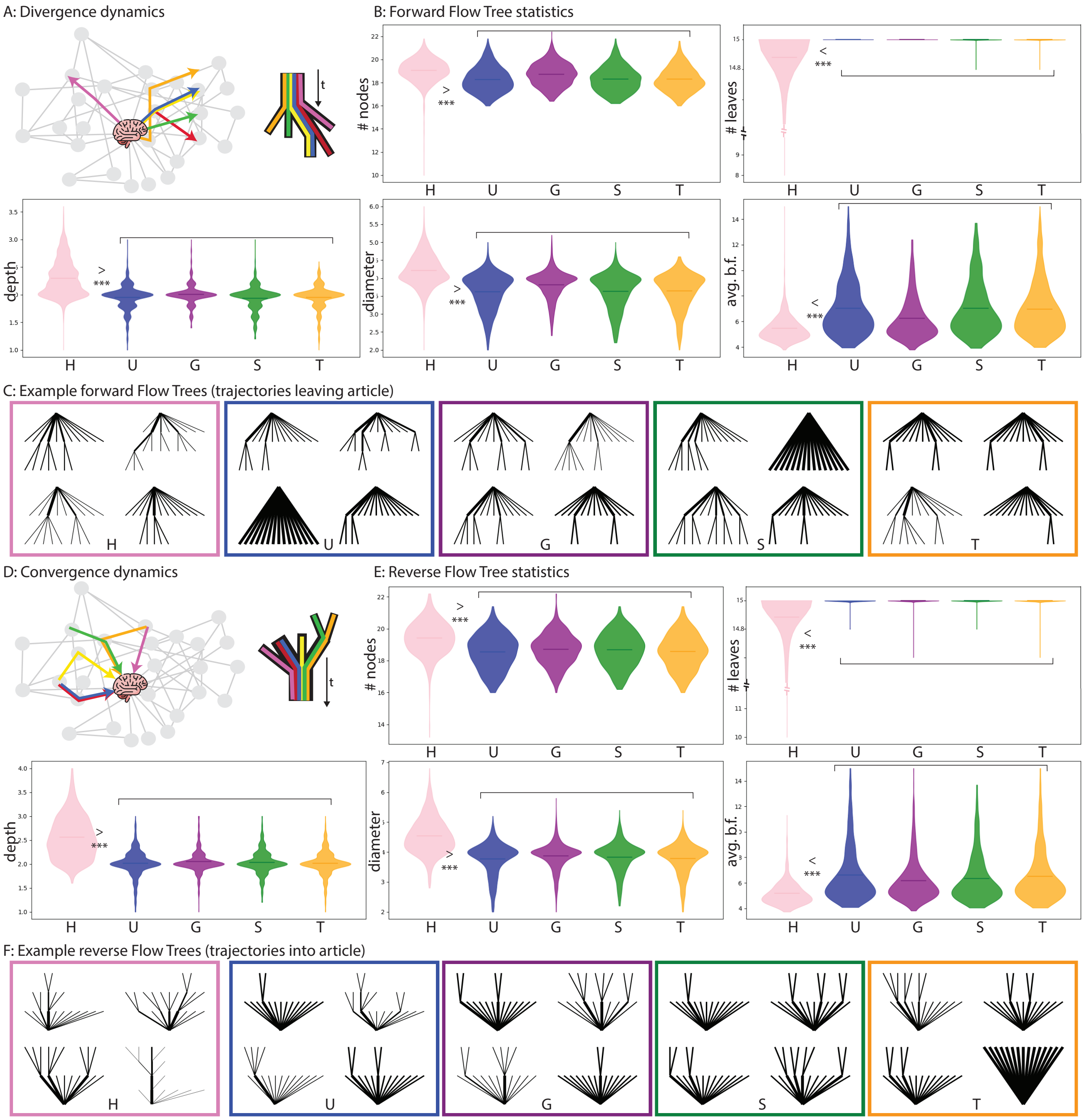}
\caption{\textbf{Convergence and divergence dynamics.} 
A, D: Forward and reverse Flow Trees encode divergence (A) and convergence (D) dynamics, respectively. 
B, E: Violin plots of five Flow Tree features for 815 of the top 1000 articles for humans (H, pink), unbiased (U, blue), graph-structure biased (G, purple), semantically biased (S, green), and targeted (T, orange) random walkers. Brackets indicate significant differences with Human (Welch's $t$-test, BH-corrected; $^{***}p<0.001$). Human Flow Trees are consistently deeper and less fan-like than all random-walker models. See Table~\ref{tab:forward_ft_stats} and Table~\ref{tab:reverse_ft_stats} for statistics. 
C, F: Example forward (C) and reverse (F) Flow Trees.
}
\label{fig:humanvsrandomft}
\end{figure*}

Although computationally convenient, traversal counts inherently collapse the temporal structure of navigation. They record that an article was visited, but elide what happened before and after. Our second analysis uses Flow Trees to preserve and compare the diverging structure of groups of trajectories leaving articles (Fig~\ref{fig:humanvsrandomft}A) and the converging structure of groups reaching them (Fig~\ref{fig:humanvsrandomft}D). 

\begin{table}[h!]
\caption{Mean values of forward Flow Tree features for humans and four random-walker models (article-level averages, $n=815$ articles). H=human, U=unbiased random walker, G=graph-structure random walker, S=semantic random walker, T=targeted random walker. Welch's $t$-test $p$-values and Cohen's $d$ are reported for each human vs. random-walker comparison.
For Cohen's $d$ effect size: small $|d|>0.2$, medium $|d|>0.5$, large $|d|>0.8$.
$^{***}$ $p<0.001$; $p$-values are corrected using the Benjamini \& Hochberg procedure.}
\centering
\begin{tabular}{lccccccccc}
\toprule
 & \multicolumn{5}{c}{Mean} & \multicolumn{4}{c}{$d$} \\
\cmidrule(lr){2-6}\cmidrule(lr){7-10}
Feature & H & U & G & S & T & \multicolumn{1}{c}{vs.\ U} & \multicolumn{1}{c}{vs. G} & \multicolumn{1}{c}{vs. S} & \multicolumn{1}{c}{vs. T}\\
\midrule
Num. nodes & \textbf{19.04} & 18.28 & 18.70 & 18.30 & 18.29 & $+0.70^{***}$ & $+0.33^{***}$ & $+0.69^{***}$ & $+0.71^{***}$ \\
Num. leaves & 14.89 & \textbf{15.00} & \textbf{15.00} & \textbf{15.00} & \textbf{15.00} & $-0.31^{***}$ & $-0.31^{***}$ & $-0.31^{***}$ & $-0.31^{***}$ \\
Depth & \textbf{2.295} & 1.946 & 2.003 & 1.938 & 1.952 & $+1.25^{***}$ & $+1.09^{***}$ & $+1.28^{***}$ & $+1.24^{***}$ \\
Diameter & \textbf{4.210} & 3.622 & 3.818 & 3.626 & 3.643 & $+1.25^{***}$ & $+0.92^{***}$ & $+1.24^{***}$ & $+1.22^{***}$ \\
Branching factor & 5.468 & \textbf{7.046} & 6.244 & 7.036 & 6.970 & $-0.96^{***}$ & $-0.57^{***}$ & $-0.95^{***}$ & $-0.93^{***}$ \\
\bottomrule
\end{tabular}
\label{tab:forward_ft_stats}
\end{table}

\begin{table}[h!]
\caption{Mean values of reverse Flow Tree features for humans and four random-walker models (article-level averages, $n=815$ articles). H=human, U=unbiased random walker, G=graph-structure random walker, S=semantic random walker, T=targeted random walker. Welch's $t$-test $p$-values and Cohen's $d$ are reported for each human vs. random-walker comparison.
For Cohen's $d$ effect size: small $|d|>0.2$, medium $|d|>0.5$, large $|d|>0.8$.
$^{***}$ $p<0.001$; $p$-values are corrected using the Benjamini \& Hochberg procedure.}
\centering
\begin{tabular}{lccccccccc}
\toprule
 & \multicolumn{5}{c}{Mean} & \multicolumn{4}{c}{$d$} \\
\cmidrule(lr){2-6}\cmidrule(lr){7-10}
Feature & H & U & G & S & T & \multicolumn{1}{c}{vs.\ U} & \multicolumn{1}{c}{vs. G} & \multicolumn{1}{c}{vs. S} & \multicolumn{1}{c}{vs. T}\\
\midrule
Num. nodes & \textbf{19.41} & 18.55 & 18.70 & 18.69 & 18.58 & $+0.79^{***}$ & $+0.69^{***}$ & $+0.67^{***}$ & $+0.77^{***}$ \\
Num. leaves & 14.88 & \textbf{15.00} & \textbf{15.00} & \textbf{15.00} & \textbf{15.00} & $-0.36^{***}$ & $-0.36^{***}$ & $-0.36^{***}$ & $-0.36^{***}$ \\
Depth & \textbf{2.560} & 2.014 & 2.047 & 2.037 & 2.017 & $+1.49^{***}$ & $+1.45^{***}$ & $+1.46^{***}$ & $+1.50^{***}$ \\
Diameter & \textbf{4.540} & 3.773 & 3.865 & 3.830 & 3.786 & $+1.34^{***}$ & $+1.27^{***}$ & $+1.28^{***}$ & $+1.35^{***}$ \\
Branching factor & 5.185 & \textbf{6.607} & 6.169 & 6.350 & 6.525 & $-0.85^{***}$ & $-0.73^{***}$ & $-0.78^{***}$ & $-0.85^{***}$ \\
\bottomrule
\end{tabular}
\label{tab:reverse_ft_stats}
\end{table}

The analysis works as follows. We consider the 1000 most frequently traversed articles, and then narrow that down to the 815 that have sufficient trajectories in both directions for all five walker types. Then, for each article, direction, and walker type we sample 15 trajectories and build a Flow Tree. We do this five times. Then, we average each of the Flow Tree features (number of nodes, number of leaves, depth, diameter, and average branching factor) across the five samples. This yields one five-dimensional feature vector that represents the dynamics of each article in each direction for each walker type (see Materials and Methods and Fig~\ref{fig:howtobuild}D-E). We compare humans to each random walker across articles using Welch's $t$-test with Benjamini-Hochberg correction and with Cohen's $d$ reported as an effect size.

Before reporting the results of the comparison, we note one potential confound. The 815 articles have higher in-degree than out-degree (mean in-degree$=96.5$, std=$122.5$; mean out-degree$=47.9$, std=$32.0$). Although the entire Wikispeedia graph is balanced by definition, the most commonly traversed articles have disproportionately high in-degree, since high in-degree means that humans have more links to click to get there, and therefore more opportunities to traverse that article. This is driven by a few very high in-degree articles, such as ``United States'' (in-degree:out-degree = 1551:294), ``United Kingdom'' (972:168), ``France'' (959:85), and ``Europe'' (933:159). To mitigate this potential bias, we repeat the Flow Tree comparison and degree regression separately on 458 of the 500 most commonly traversed articles (mean in-degree$=132.4$, std=$151.5$; mean out-degree$=53.4$, std=$35.9$), which have an even higher skew, as well as on 357 of the 501st-1000th most traversed articles (mean in-degree$=50.4$, std=$32.5$; mean out-degree$=41.0$, std=$24.6$), which are much more balanced, and find the results qualitatively the same (see~\nameref{sec:robustness}). All results reported below pertain to the full set of 815 articles.

\subsubsection*{Forward Flow Trees}
Forward Flow Trees encode how trajectories diverge, that is, how different trajectories depart from a particular article of interest~(Fig~\ref{fig:humanvsrandomft}A). Human Flow Trees are deeper, have larger diameter, and have lower average branching factor than those of all random-walker models, with large effect sizes (vs.\ unbiased: depth $d=1.25$, diameter $d=1.25$, branching factor $d=-0.96$; vs.\ graph-structure: depth $d=1.09$, diameter $d=0.92$, branching factor $d=-0.57$; see Fig~\ref{fig:humanvsrandomft}B and Table~\ref{tab:forward_ft_stats}). This indicates that humans do not scatter directly after leaving an article; instead they tend to follow a common sequence of steps, resulting in Flow Trees that extend, rather than branch immediately. These human Flow Trees are deeper and less fan-like than those that the random-walker null models generate (see Fig~\ref{fig:humanvsrandomft}C), though trunk-like divergence need not imply that humans have multi-step shared structural patterns. 

The semantic and the targeted random walkers generate Flow Trees nearly identical to those of the unbiased random walker across all features, and the graph-structure random walker is consistently the closest to humans. This may appear at first to contradict Results Section~(1), where the semantic random walker best matched humans on word frequency, but the two statistics measure different things. Traversal counts measure individual articles, while Flow Trees measure sequences of them (and branching factor measures how concentrated choices are at each step).

The ways that the random walkers ``decide'' their next-click explains this. Word frequency varies relatively little among the neighbors of any given article, so the resulting distribution over neighboring articles is close to uniform. This makes it so the random walker is biased as to which articles it visits overall, but not in how a group spreads out from any one article. Its Flow Trees, therefore, are nearly indistinguishable from those of the unbiased random walker. The graph-structure random walker, in contrast, is more selective, concentrating probability on its high-degree neighbors of each article, so groups of its trajectories stay together for longer before splitting. This results in deeper and less fan-like Flow Trees.

This makes the graph-structure random walker the closest of the four to humans in Flow Tree shape, but this does not necessarily mean that specifically weighting by degree is the closest thing to what humans are doing. Concentration of choice at each article and agreement about which articles to route through are different things. Results Sections (3)-(6) tease these two apart.

Humans produce slightly fewer distinct trajectories on average than the random walkers, as indicated by a smaller average number of Flow Tree leaves (num. leaves: $d \approx -0.31$ against all random-walker types, $p<0.001$). All four random-walker models produce a nearly identical mean number of leaves (mean$=15.00$), which is also the number of distinct trajectories each Flow Tree is constructed from, while humans occasionally deviate from this baseline (mean$=14.89$), which means that at least two human trajectories were sometimes decision-for-decision identical over their recorded length. 

More striking than the difference in means is the difference in variance among articles. All random walkers have zero variance in number of leaves, sitting at the 15-leaf ceiling, while humans vary across articles ($\sigma^2 = 0.264$; see the top right panel of Fig~\ref{fig:humanvsrandomft}B). The variance in humans suggests that some articles produce highly cohesive outgoing trajectories, while others scatter players across many distinct paths. This variation in how cohesively humans depart from different articles is the first signature of collective behavior in the data. Humans respond to article-specific properties in a coordinated way that no null model captures. We characterize it in Results Sections~(3)-(5).

\subsubsection*{Reverse Flow Trees}
Reverse Flow Trees encode how trajectories converge, that is, how different trajectories end up at a particular article of interest~(Fig~\ref{fig:humanvsrandomft}D). Their statistics tell a similar story to forward Flow Trees, where the same qualitative pattern holds: human reverse Flow Trees are deeper, have larger diameter, and have lower branching factor than all random walkers (see Fig~\ref{fig:humanvsrandomft}E and Table~\ref{tab:reverse_ft_stats}). Notably, the effect sizes are larger for depth and diameter (but not branching factor) in the reverse direction than in the forward direction (depth: $d=1.49$ vs.\ unbiased, compared to $d=1.25$ forward; diameter: $d=1.34$ vs.\ $d=1.25$ forward; branching factor: $d=-0.85$ vs.\ $d=-0.96$ forward; Table~\ref{tab:reverse_ft_stats}). The graph-structure random walker, which was the closest to humans for forward Flow Trees, is also the closest to humans in the reverse direction. Groups of trajectories coming to an article are more sequentially structured for humans than for the null models. This results in deeper Flow Trees (see Fig~\ref{fig:humanvsrandomft}F). This implies that humans converge on articles in a more cohesive manner than graph structure alone predicts.

Humans also produce slightly fewer distinct trajectories in the reverse direction (see Fig~\ref{fig:humanvsrandomft}E, top right panel). The random walkers again sit at the 15-leaf ceiling ($\sigma^2_{rw} < 0.001$), while the humans vary ($\sigma^2=0.206$). This confirms that there is article-level heterogeneity in human behavior in both convergence and divergence. 

Human Flow Trees clearly differ from those of random walkers, but they are partially driven by the same underlying topological constraints. Wikispeedia articles with many outgoing links give navigators a large decision space, which produces simpler, more fan-like Flow Trees for humans and random walkers alike. Likewise, articles that have many incoming links attract traffic from many different directions, leading to shallower, fan-like trees. In the next section, we investigate how much out-degree and in-degree drive Flow Tree shape in humans, and in random walkers.

\subsection*{(3) Degree explains some, but not all, of human Flow Tree shape}

Our third analysis seeks to quantify the relationship between an article's connectivity in the underlying Wikispeedia graph and the shape of its Flow Trees. The number of links leaving an article constrains a navigator's hypothesis space; the more outgoing links, the less likely that different outgoing trajectories share subsequent decisions. High-out-degree articles should therefore produce shallower, more fan-like Flow Trees, while low-out-degree articles force decisions through a bottleneck, producing deeper and more structured trees. The same logic applies to in-degree and reverse Flow Trees.

\begin{figure}[!ht]
\centering 
\includegraphics[width=\textwidth, page=6]{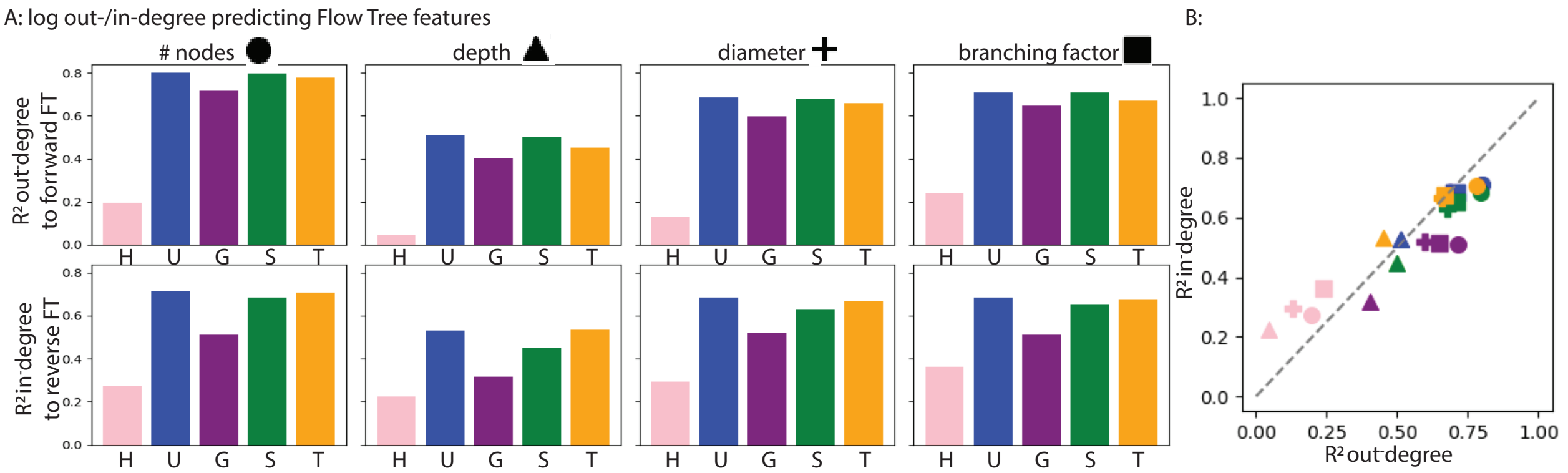}
\caption{\textbf{Regression.} A: $R^2$ of (top row) log out-degree predicting forward Flow Tree features and (bottom row) log in-degree predicting reverse Flow Tree features for humans (H, pink), unbiased (U, blue), graph-structure (G, purple), semantic (S, green), and targeted (T, orange) random walkers across four features (columns, number of nodes is circle; depth is triangle; diameter is cross; average branching factor is square). Here we plot the univariate $R^2$; bivariate values appear in Table~\ref{tab:regression}. Human $R^2$ is lower than that of all random-walker models in all directions. B: Scatter plot of reverse in-degree $R^2$ (y-axis) vs. forward out-degree $R^2$ (x-axis) for each walker-type (color) feature (shape) combination. Points above the dashed line (identity) are those for which in-degree predicts reverse Flow Tree features better than out-degree predicts forward Flow Tree features. This is true for all human points.}
\label{fig:regression}
\end{figure}

We test this by regressing each Flow Tree feature against log out-degree (forward Flow Trees) and log in-degree (reverse Flow Trees), separately for each walker type across the same 815 articles (Fig~\ref{fig:regression}A, Table~\ref{tab:regression}). For each feature and direction we fit both the univariate model containing only the relevant degree, and a bivariate model with both degrees, so we can ask whether the other carries any additional information. We omit number of leaves, because its variance for random walkers is zero.

\begin{table}[h!]
\centering
\caption{Degree regression summary for forward and reverse Flow Trees
($n = 815$ articles).
$R^2_{\mathrm{dom}}$: univariate $R^2$ of the dominant-direction degree
predictor (out-degree for forward, in-degree for reverse).
$R^2_{\mathrm{both}}$: multivariate $R^2$ with both degrees included.
Adding the opposite-direction degree contributes negligibly for all walker
types ($\Delta R^2 \leq 0.006$).
Number of leaves is also omitted, since random-walker variance is near zero (all trajectories produce the maximum of 15 distinct paths),
making degree-based prediction uninformative.
H=human, U=unbiased random walker, G=graph-structure random walker, S=semantic random walker, T=targeted random walker.}
\centering
\begin{tabular}{lcccccccccc}
\toprule
 & \multicolumn{5}{c}{$R^2_{\mathrm{dom}}$} & \multicolumn{5}{c}{$R^2_{\mathrm{both}}$} \\
\cmidrule(lr){2-6}\cmidrule(lr){7-11}
Feature & H & U & G & S & T & H & U & G & S & T \\
\midrule
\textit{Forward (out-deg.)} \\
\hspace{0.2cm}Num. nodes & 0.195 & 0.798 & 0.715 & 0.795 & 0.779 & 0.199 & 0.799 & 0.715 & 0.795 & 0.779 \\
\hspace{0.2cm}Depth & 0.047 & 0.510 & 0.403 & 0.500 & 0.452 & 0.052 & 0.511 & 0.403 & 0.501 & 0.454 \\
\hspace{0.2cm}Diameter & 0.131 & 0.684 & 0.599 & 0.676 & 0.657 & 0.135 & 0.685 & 0.599 & 0.677 & 0.660 \\
\hspace{0.2cm}Branching factor & 0.241 & 0.709 & 0.648 & 0.708 & 0.670 & 0.246 & 0.710 & 0.648 & 0.709 & 0.672 \\
\midrule
\textit{Reverse (in-deg.)} \\
\hspace{0.2cm}Num. nodes & 0.272 & 0.712 & 0.511 & 0.685 & 0.706 & 0.274 & 0.713 & 0.512 & 0.687 & 0.708 \\
\hspace{0.2cm}Depth & 0.223 & 0.528 & 0.316 & 0.449 & 0.533 & 0.223 & 0.529 & 0.317 & 0.450 & 0.533 \\
\hspace{0.2cm}Diameter & 0.294 & 0.682 & 0.517 & 0.629 & 0.668 & 0.294 & 0.684 & 0.520 & 0.632 & 0.670 \\
\hspace{0.2cm}Branching factor & 0.362 & 0.685 & 0.513 & 0.651 & 0.676 & 0.368 & 0.689 & 0.518 & 0.657 & 0.679 \\
\bottomrule
\end{tabular}
\label{tab:regression}
\end{table}

\paragraph{Out-degree predicts forward Flow Trees, in-degree predicts reverse Flow Trees, and the opposite-direction degree adds nothing.}
Intuitively, the number of options there are for departure should impact the cohesiveness of divergence, the number of options there are for arrival should impact the cohesiveness of convergence, and neither should carry much information about the other direction. And, indeed, for every walker type and every feature, the matched-direction degree is the dominant predictor. Adding the opposite-direction degree to the model contributes almost nothing: the bivariate $R^2$ exceeds the dominant-direction univariate $R^2$ in every row of Table~\ref{tab:regression} by at most $0.006$. This holds for both humans and the random walkers, allowing us to treat the forward and reverse residuals in Results Section~(4) as separate axes.

\paragraph{Degree explains less of the variance in human Flow Trees than in those of any random walker.} 
Humans follow the same direction-specific pattern, but with substantially lower $R^2$ across every feature and in both directions (Fig~\ref{fig:regression}A). The gap is largest for depth, where out-degree explains $4.7$\% of the variance in human forward Flow Trees compared to $51.0$\% for the unbiased random walker, and for diameter ($13.1$\% human vs.\ $68.4$\% unbiased random walker). Even for branching factor--the feature which degree explains the most human variance, making this the most conservative comparison--degree explains $24.1$\% of human variance compared to $70.9$\% for the unbiased random walker. 
To test these differences, we used bootstrapping (with 10,000 resamples, recomputing all $R^2$ values on each resample), and find that the human $R^2$ is below that of every random walker for every feature in both directions (all 95\% CIs on $\Delta R^2$ exclude zero, and the smallest $\Delta R^2=0.094$, 95\% CI $[0.013,0.168]$ for reverse Flow Tree depth compared to the graph-structure random walker.) Random walkers make decisions using only local graph structure, so their Flow Trees are tightly determined by it. Humans are drawing on additional signals, and the shortfall in $R^2$ is a direct measure of how much of their navigation graph structure fails to capture.

Notably, the semantic and targeted random walkers are nearly indistinguishable from the unbiased random walker throughout, so the human shortfall is not reproduced by weighting decisions toward semantically prominent or target-similar articles. The graph-structure random walker has consistently the lowest $R^2$ of the four null models, since degree-weighting concentrates its trajectories on a subset of neighbors and thereby loosens the relationship between an article's raw degree and the resulting tree shape.

\paragraph{In-degree predicts reverse Flow Trees better than out-degree predicts forward Flow Trees for humans.}
Comparing the two directions reveals a dissociation (Fig~\ref{fig:regression}B). For twelve of the sixteen random-walker-feature combinations, forward $R^2$ exceeds reverse $R^2$. For humans, the opposite holds on every feature--reverse $R^2$ exceeds forward $R^2$ (depth: $0.223$ vs.\ $0.047$; diameter: $0.294$ vs.\ $0.131$; branching factor: $0.362$ vs.\ $0.241$)--and all four human points fall above the line, against four of sixteen random-walker points. Part of this pattern depends on which articles are analyzed (see~\nameref{sec:robustness}). For number of nodes, depth, and diameter, it holds both in the 500 most-traversed articles and in the more degree-balanced articles ranked 501–1000. For branching factor, it holds only among the most-traversed articles and vanishes in the balanced subset (reverse $0.149$ vs. forward $0.155$). One reading of this asymmetry is that the number of outgoing links an article has places a weak constraint on where human players go, whereas the number of incoming links constrains more tightly where human players are coming from. 

This asymmetry is not just a fact pertaining to human navigation. Forward and reverse traversal are formally distinct processes on a directed graph, regardless of the entity that is traversing it (see~\nameref{sec:forward-reverse-dynamics}).

Degree accounts for a large share of random-walker Flow Tree shape and a much smaller share of human shape, leaving roughly three quarters of the variance in human forward branching factor unexplained. In the next results section we look into this unexplained variance to define a two-dimensional description of an article's navigational type.

\subsection*{(4) Degree residuals define article-specific navigational types}

Our fourth analysis asks whether there is structure to the variance in Flow Trees that degree fails to explain. We use the residuals from the forward and reverse regressions above as coordinates describing how navigation dynamics differ in departure and arrival from what structural connectivity alone would predict. We use average branching factor because it is the feature on which degree explains the most human variance ($24.1$\% forward, $36.2$\% reverse), so it is the most conservative choice. 
A residual allows us to compare an article to other articles with the same degree, rather than to the degree itself. A positive forward residual means that an article scatters departing navigators more widely than is typical for an article with that out-degree; a positive reverse residual means that it receives arriving navigators from a more diverse set of routes than its in-degree predicts. Because branching factor averages over all splits along the route, not just at the first click, the residual reflects how players route through the following clicks, not only on which link they click next.

\subsubsection*{Degree residuals carry a reproducible, article-specific signal along two independent axes}
First, we split each article's human trajectories into two disjoint halves, build Flow Trees and the corresponding residuals from each half independently, and correlate the two across articles (there are 563 articles with at least 30 trajectories in each direction; see Materials and Methods). We find the residuals partially reproducible (forward $r=0.20$, $[0.16, 0.22]$; reverse $r=0.37$, $[0.33, 0.44]$;
when trajectory pools are matched to the random walkers, this becomes $r=0.13$ and $0.30$, Results Section~(5)). The residuals carry an article-specific signal. The forward and reverse residuals are only weakly correlated with each other ($r=0.04$, $p=0.254$), so they do span a two-dimensional space. The human forward residual is unrelated to that of any random walker ($|r| \le 0.05$, all $p>0.15$). The reverse residual is weakly shared with all four ($r=0.16$–$0.21$, all $p<0.001$), so a small part of it reflects the Wikispeedia graph. In-degree counts an article's incoming links but not how much traffic each one carries, and convergence depends on both (see~\nameref{sec:forward-reverse-dynamics}).

\subsubsection*{Four article types}
\begin{figure}[!ht]
\centering 
\includegraphics[width=\textwidth, page=7]{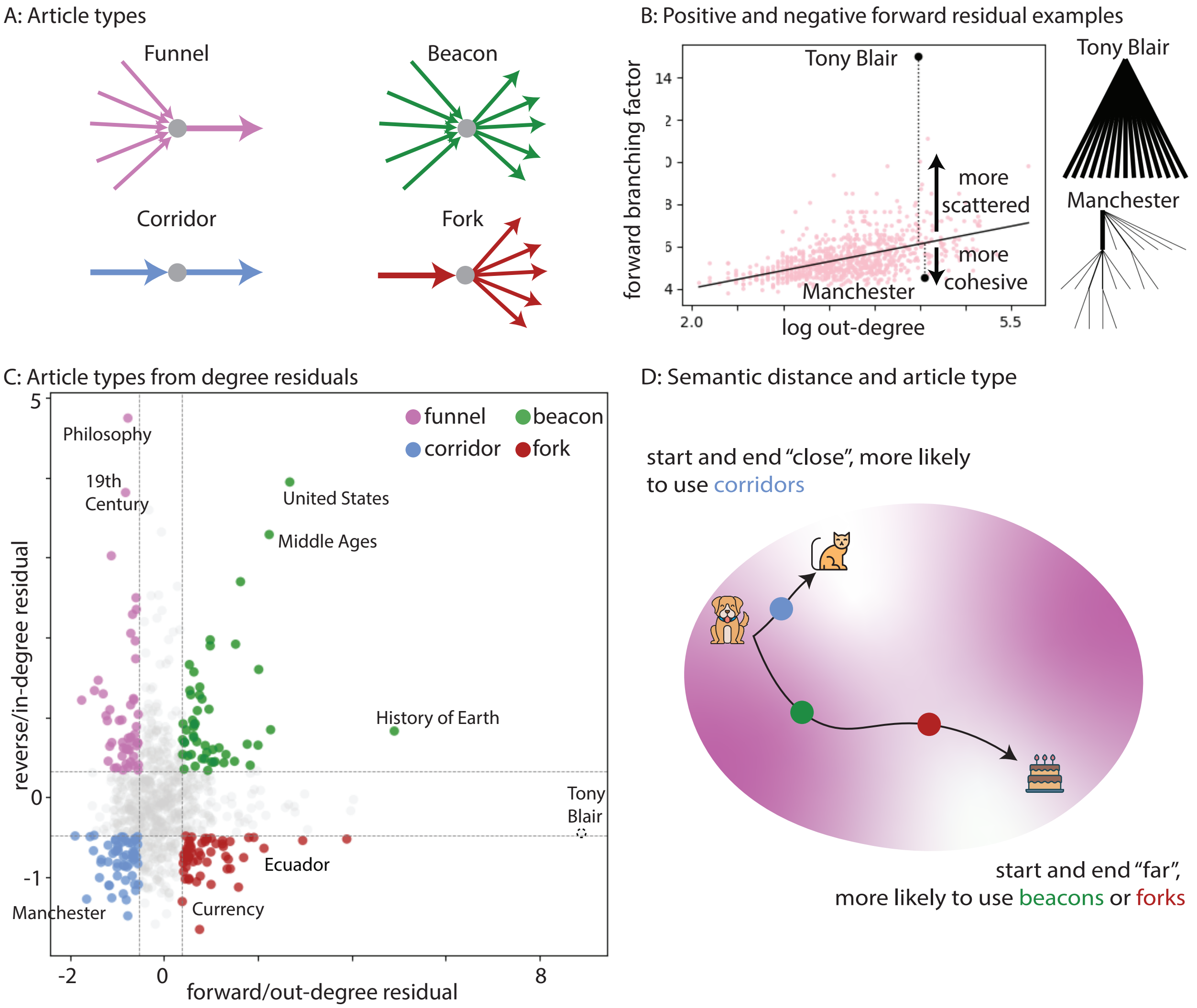}
\caption{\textbf{Two-dimensional space of article functional types.} 
A: Four article types. Funnels (pink) have diverse arrival and cohesive departure. Beacons (green) scatter in both directions. Corridors (blue) channel in both directions. Forks (red) have cohesive arrival, diverse departure. 
B: Shown is a plot of forward Flow Tree branching factor vs. log out-degree, with a light pink dot for each of the 815 articles. The regression line is shown in black. An article's residual is the vertical distance from its point to the degree regression line. ``Tony Blair'' has positive forward residual and scatters departing humans far more than its out-degree predicts. Its forward Flow Tree is a fan. ``Manchester'' has negative forward residual and channels them more than predicted, resulting in a more structured forward Flow Tree.
C: Each point is one of 815 articles, positioned by its forward and reverse branching-factor residual. Both axes are degree-free by construction and near-independent ($r=0.040$). Dashed lines mark the upper and lower quartiles of each axis, defining the four corner categories. Categories do not differ systematically in out-degree or semantic centrality, though they differ in in-degree. A few example articles are labeled on the graph. 
D: Trajectories that are between more semantically distant start and target articles (e.g., dog and cake) more often pass through beacons and forks. Trajectories whose start and target are closer (e.g., dog and cat) more often pass through corridors.}
\label{fig:typology}
\end{figure}

Splitting articles at the upper and lower quartiles of each residual axis gives four types defined by human behavior, one in each corner. These are based on the Cartesian product of \{diverse vs. few origins\} and \{scatter vs. channel\} (see Fig~\ref{fig:typology}A and Table~\ref{tab:typology}):
\begin{itemize}
    \item \textbf{Funnels} ($n=49$) collect navigators arriving from diverse origins and channel them forward cohesively.
    \item \textbf{Beacons} ($n=51$) collect navigators arriving from diverse origins and scatter them widely.
    \item \textbf{Corridors} ($n=56$) collect navigators from a narrow set of routes and channel them forward cohesively.
    \item \textbf{Forks} ($n=54$) collect navigators arriving from a narrow set of routes and then disperse them widely.
\end{itemize}
Articles that do not belong to any of the above quartile combinations are unclassified ($n=605$). The four types describe distinct navigational roles.  
Articles have previously been classified by comparing the clicks they receive with the clicks they send on~\cite{gildersleve2018inspiration}. Our types are defined by how groups of navigators split over several clicks, rather than by single-step traffic, and they are defined relative to what an article's degree predicts, so they describe how an article is used rather than how well it is connected.
Funnels, such as ``Philosophy'' and ``World War II'', draw navigators in from many directions and then release them along a shared route.
Beacons, such as ``History of Earth'' and ``United States'', are broad, general-purpose articles that serve as crossroads.
Forks, such as ``Animal rights'' and ``Ecuador'', are reached along few routes, but scatter departing navigators.
Corridors, such as ``List of elements by name'' and ``Victoria of the United Kingdom'', channel navigators in both directions, and are typically reached and left via one or two canonical routes.

\begin{table}[h!]
\centering
\caption{Four navigational article types defined by forward and reverse branching-factor residuals. Wikispeedia articles are assigned to a type if they are in the upper or lower quartile of both residual axes; all other articles are unclassified ($n=605$). In-deg.\ and out-deg.\ are raw means, shown for reference only. Residuals already remove each degree's effect, so a type reflects how an article is used relative to what its degree predicts, not how connected it is. We provide the five most extreme examples based on Euclidean distance from the origin. F=forward, R=reverse, WF=log word frequency.}
\small
\setlength{\tabcolsep}{4pt}
\begin{tabular}{llccccccp{4cm}}
\toprule
Type & Dynamics & $n$ & F res. & R res. & In-deg. & Out-deg. & WF & Example articles \\
\midrule
Funnel & diverse in, cohesive out & 49 & $-0.85$ & $+1.11$ & 116 & 67 & 4.41 & Philosophy, 19th century, Time zone, World War II, Uganda \\
Beacon & diverse in, diverse out & 51 & $+1.06$ & $+1.00$ & 136 & 59 & 4.27 & United States, History of Earth, Middle Ages, Time, Somalia \\
Corridor & cohesive in, cohesive out & 56 & $-0.92$ & $-0.79$ & 89 & 53 & 4.46 & Manchester, Henry VIII of England, List of elements by name, 17th century, Victoria of the United Kingdom \\
Fork & cohesive in, diverse out & 54 & $+0.96$ & $-0.74$ & 178 & 58 & 4.47 & Spain, Animal rights, Ecuador, Philippines, City \\
\bottomrule
\end{tabular}
\label{tab:typology}
\end{table}

\subsubsection*{Article types are not distinguished by structural or semantic centrality}
It is possible that these four types could be a mere relabeling of that which we considered in Results Section (1), that is, how well connected (graph centrality measures) or how common (semantic centrality) the articles are. They are not. Mean log out-degree is nearly constant across the types (beacon 3.89, funnel 3.99, fork 3.94, corridor 3.92; one-way ANOVA $F=0.27$, $p=0.85$), as is mean log word frequency (beacon 4.27, funnel 4.41, fork 4.47, corridor 4.46; $F=0.92$, $p=0.43$). Semantic centrality explains under 1\% of residual variance in either direction (log word frequency vs. forward residual: $r=-0.064$, $R^2=0.004$; vs. reverse residual: $r=0.016$, $R^2<0.001$).

Log in-degree is the one exception. It varies across the four types (beacon 4.35, funnel 4.13, fork 4.80, corridor 4.37; $F=5.33$, $p=0.001$), and three of six pairwise comparisons survive Benjamini-Hochberg correction for multiple tests (fork vs. funnel: $p=0.004$, $d=+0.67$; fork vs. corridor: $p=0.004$, $d=+0.63$; fork vs. beacon: $p=0.025$, $d=
+0.50$).
The ordering runs counter to raw connectivity; forks have the highest mean log in-degree of the four types, yet their reverse residual is strongly negative. In other words, the most-connected type according to in-degree shows among the most cohesive, least diverse pattern of convergence, the opposite of what degree alone would predict.
This is what a coordinate system based on usage rather than structure should produce. Beacon, funnel, and corridor do not differ significantly from one another in log in-degree (all $p>0.27$), so the dissociation is specific to forks. Although each residual is orthogonal to its respective degree by construction, in- and out-degree are themselves correlated across articles, so it is possible that a small amount of cross-degree signal could survive. And indeed, a very small amount does.
Log in-degree explains $0.55\%$ of the variance in the forward residual ($r=0.074$, $p=0.034$) and log out-degree $0.73\%$ of the reverse residual ($r=0.085$, $p=0.015$). We therefore cannot claim that these types are completely free of the influence of degree, but rather that structural connectivity and word frequency account for very little of what separates them; what does separate them is their manner of use.

\subsubsection*{Beacon and fork use is associated with the semantic distance between start and target} 
If beacons function as zoom-out points, they should be used most when the start and target are conceptually far apart. Forks could also serve as zoom-out points, as they also widely disperse navigators on departure. We test these hypotheses by classifying each successful trajectory according to which article types it passes through, and comparing the semantic distance between start and target articles ($\delta = 1-s$, using the same language-model-based computation we use for the targeted random walker, see~\nameref{sec:matmeth-rw} in Materials and Methods) across the resulting groups. 

Trajectories that pass through a beacon or a fork have start and target articles that are semantically further apart than trajectories that do not (beacon: $\delta = 0.446$ vs.\ $0.424$, $d = 0.15$; fork: $0.450$ vs.\ $0.423$, $d = 0.19$). Trajectories that pass through a corridor have start and target articles that are semantically closer together ($0.423$ vs.\ $0.433$, $d = -0.07$). And trajectories that pass through a funnel show no meaningful difference ($0.430$ vs.\ $0.432$, $d = -0.03$).

Trajectories that pass through any of the four article types are longer on average, and longer trajectories have more opportunities to pass through any fixed set of articles. All four effects remain after controlling for trajectory length (ordinary least squares with trajectory length as a covariate; beacon $t = 14.0$, fork $t = 17.1$, corridor $t = -7.8$, all $p<0.001$; funnel $t = -2.9$, $p = 0.003$). These effects are small and correlational, and the causal direction could run either way. Articles whose departing players scatter more than their out-degree predicts may be preferentially used when the start and target articles are semantically far apart. Alternatively, routing semantically distant start–target pairs through an article may itself be what scatters the players who depart from it.


\subsection*{(5) Article types are reproducible for humans and not recovered by any random walker}
We repeat the split-half procedure for each random walker, matching the number of trajectories available across navigators (described in Results Section~(4)). In the forward direction, human residuals ($r=0.13$ $[0.07, 0.21]$) are more reproducible than those of the unbiased random walker ($r=-0.01$ $[-0.08, 0.09]$; significantly higher in four of five splits) and comparable to the semantic ($r=0.04$ $[-0.02, 0.14]$), targeted ($r=0.04$ $[0.02, 0.05]$), and graph-structure ($r=0.12$ $[0.09, 0.22]$) random walkers. In the reverse direction every navigator is more reproducible than in the forward direction (unbiased $r=0.19$ $[0.16, 0.22]$, semantic $r=0.23$ $[0.21, 0.27]$, targeted $r=0.23$ $[0.18, 0.27]$, graph-structure $r=0.34$ $[0.26, 0.41]$, human $r=0.30$ $[0.25, 0.35]$), and humans are not consistently more reproducible than any random walker. Thus, reproducibility alone does not distinguish humans.

What does distinguish them is which articles are labeled with which type. Classifying articles from two disjoint halves of the human data yields types that agree above chance (Cohen's $\kappa=0.36$ $[0.22, 0.49]$ across five splits; $53\%$ agreement against $25\%$ chance). Applying the same classification to each random walker's residuals yields types that agree with the human ones only at levels predicted by chance (graph-structure: $\kappa=0.00$, 95\% CI $[-0.13, 0.14]$, $26.7\%$ agreement among the 60 articles typed by both; unbiased, semantic, and targeted: $\kappa=0.06$-$0.09$, all binomial $p>0.1$). For the graph-structure random walker, the articles typed at all overlap with humans no more than chance would predict ($\chi^2(1, N=815)=0.5$, $p=0.49$). Human navigational types are thus noisy but reproducible, and they are not the types that graph structure or word frequency would assign.

Because every walker traverses the identical directed graph from the identical starting articles, this difference cannot be attributed to graph topology alone. We interpret it as tacit agreement among players, who never communicate, about how particular articles are used.

\subsection*{(6) First-order click preferences are part, but not all, of collective knowledge}
Results Sections~(1)-(5) establish that human navigation differs from models of graph structure or language statistics. This allows us to ask a new question about this leftover structure. How much of this collective knowledge is based on individual clicks versus how much is based on the overall trajectory?

\begin{figure}[!ht]
\centering 
\includegraphics[width=\textwidth, page=9]{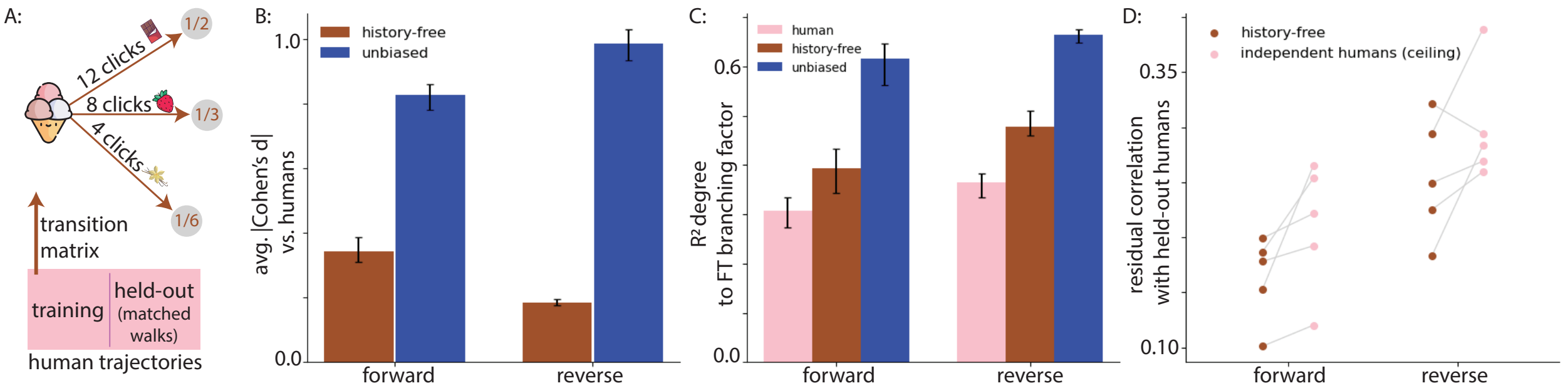}
\caption{\textbf{The history-free human model partially reproduces human Flow Tree shape but differs significantly from human navigation.}
A: The history-free human model chooses the next article according to human click counts at the current article. The human trajectories are randomly split in half: half are used to generate the transition matrix, and the other half are then used as reference to generate matched random walks and generate the corresponding Flow Trees.
B: Average Cohen's $d$ effect size between humans and the history-free human model (brown) and between humans and the unbiased random walker (blue), for forward and reverse.
C: Variance in average branching factor explained by log degree (out-degree for forward and in-degree for reverse) for humans, the history-free human model, and the unbiased random walker. The ordering of human $<$ history-free $<$ unbiased holds for every random split.
D: Correlation of each article's degree residual between held-out humans and either the history-free human model (brown) or the other half of human data (pink). The latter acts as a ceiling. In B and C, the colored bars show averages and the error bars show the range over the five random splits.}
\label{fig:history-free}
\end{figure}

To probe this question, we use the history-free human model (see Fig~\ref{fig:history-free}A), which traverses the graph based just on human first-order click preferences at each article with no memory of previous articles or knowledge of the target (see~\nameref{sec:matmeth-hf} in Materials and Methods). 
If this history-free human model and humans are identical, then this likely means human navigation on Wikispeedia is a sequence of local decisions that just depend on the current article. Differences between the two indicate temporally extended structure in human navigation.

We first repeat the analysis in Results Section (2), analyzing Flow Tree features. The history-free human model, which biases each decision according to how often humans clicked each link from the current article, produces Flow Trees closer to human Flow Trees than do any of the four random walkers (see Fig~\ref{fig:history-free}B); the mean absolute effect size across the five features is $d=0.34$ $[0.31, 0.39]$ forward and $d=0.19$ $[0.18, 0.20]$ reverse, compared to $d=0.83$ $[0.78, 0.86]$ and $d=0.99$ $[0.93, 1.03]$ for the unbiased random walker. Its Flow Trees nonetheless differ significantly from human Flow Trees on all five forward features in every split, and on number of leaves, depth, and diameter for reverse Flow Trees in every split (Welch's $t$-test, BH-corrected), but never on reverse branching factor. In the reverse direction, per-article click preferences thus reproduce how widely groups fan out at each step, but not how long they stay together.

The next two analyses show what single-click preferences miss. We repeat the degree regression analysis done in Results Section (3). Degree explains more of the variance in branching factor for the history-free human model than for humans, but less than for the unbiased random walker (forward $R^2$: human $0.31$ $[0.27, 0.33]$ $<$ history-free $0.39$ $[0.34, 0.43]$ $<$ unbiased $0.62$ $[0.56, 0.65]$; reverse: $0.37$ $[0.33, 0.38]$ $<$ $0.48$ $[0.46, 0.51]$ $<$ $0.66$ $[0.65, 0.67]$; computed on held-out data so differ from Table~\ref{tab:regression}; this ordering holds in all five splits; see Fig~\ref{fig:history-free}C). Degree explains less of navigation dynamics as the model becomes more human-like. It explains the most for the unbiased random walker, then less for the history-free human model, and then the least for humans. First-order click preferences account for part of the departure of human navigation from graph structure, and the remaining difference is significant in every random split.

We then correlate the degree residuals of each null model with human residuals across articles (see Fig~\ref{fig:history-free}D). Residuals from the history-free human model correlate with human residuals more than do residuals from the unbiased random walker (forward: $r=0.14$ $[0.11, 0.19]$ vs. $0.05$ $[-0.03, 0.15]$; reverse: $0.25$ $[0.18, 0.32]$ vs. $0.06$ $[0.03, 0.10]$). Since human residuals themselves contain sampling noise, we compare these correlations against the correlation between two independent groups of humans, the held-out trajectories and the trajectories used to train the history-free human model (see Fig~\ref{fig:history-free}A). The history-free human model falls below this human-human ceiling in every split in the forward direction ($r=0.16$ $[0.10, 0.20]$ vs. $0.21$ $[0.12, 0.26]$; 451-482 articles), and four of five splits in the reverse direction ($0.25$ $[0.18, 0.32]$ vs. $0.30$ $[0.26, 0.39]$; 582-612 articles). First-order click preferences recover part of which articles humans use differently from what their degree predicts, but in the forward direction they consistently fall short of what a second group of humans shares.

\section*{Discussion}
We find that human navigation through Wikispeedia is constrained relative to random walks. Groups of players leaving an article often share several clicks before diverging, and groups arriving at one share the clicks that brought them there. This results in human Flow Trees that are consistently deeper and less fan-like than those of four different random-walker null models.

Consistent structure emerges when we consider navigation as a collection of trajectories, rather than individual trajectories on their own. While ant trails~\cite{vittori2006path} and dirt paths across lawns~\cite{helbing2001self} result through so-called ``stigmergy'', whereby each trajectory leaves a trace in the environment upon which structure and consistency self-reinforce, humans in Wikispeedia leave no explicit trace for others to follow. Their agreement must come from shared knowledge or strategies that players bring with them to the game. Our null models let us decompose such agreement. Navigation based on graph structure alone results in fan-like Flow Trees whose shape is largely determined by degree; shared single-click preferences account for part of the human departure from that baseline; and a significant remainder extends across clicks.

\subsection*{Collective knowledge in Wikispeedia}
Every navigator in our study traverses the same directed graph from the same starting articles, and their trajectories are truncated to the same length for analysis. This allows us to consider differences between navigators. For the unbiased random walker, an article's degree explains about 70\% of the variance in branching factor; for humans, it explains a quarter in the forward direction and a third in the reverse (Results Section~(3)). The history-free human model, which follows human click frequencies at each article but has no memory or target, closes part of this gap in Flow Tree shape. On held-out data, degree explains 62\% of the variance in forward branching factor for the unbiased random walker, 39\% for the history-free human model, and 31\% for humans (and for reverse, 66\%, 48\%, and 37\%, respectively), and this ordering holds in every split. Flow Trees constructed from history-free human model trajectories are also closer to those constructed from human trajectories than are those from unbiased random walker trajectories (mean $|d| = 0.34$ vs.\ $0.83$ forward; $0.19$ vs.\ $0.99$ reverse), yet they still differ significantly from human Flow Trees on most features in every split.

Part of the shared structure in human navigation over Wikispeedia is in agreement about the next link to click. At a given article, human choices are more concentrated than those of any random walker, in ways that neither degree, word frequency, nor title similarity to the target predicts. According to Hecker's taxonomy~\cite{hecker2012knowledge}, this is collective knowledge as the intersection of individual knowledge--overlapping expectations about how different concepts relate. The remainder, which the history-free human model misses, requires players' choices to depend on where they have been or where they are going.

Human Flow Trees are deeper, have larger diameter, and are less fan-like than random-walker Flow Trees (Results Section~(2)). One possible conclusion is that these longer, more structured trunks indicate that humans share planned, multi-click routes. The history-free human model shows that part of this shape is attainable with no temporally extended plan. If most humans at an article choose the same next article, and then agree again, groups can stay together for several clicks without the need for memory. But the history-free human model does not reproduce how long groups stay together (depth and diameter differ in every split, in forward and reverse). It also does not reproduce how far human Flow Trees depart from what degree predicts. And, for divergence, it does not reproduce which articles depart from their degree as consistently as they do for a second group of humans. Together, these differences are our evidence for temporally extended structure.

This relates to a question that order-selection methods have struggled to settle. Singer et al.\ found that after penalizing model complexity, a memoryless model was preferred for page-level navigation in a related Wikipedia game, while Wikispeedia navigation between topics was best described by a second-order chain~\cite{singer2014detecting}. They attributed the page-level result in part to data sparsity. Arora et al.\ found that in free Wikipedia browsing, where a reader arrives from says little about where they go next for most articles, but a small fraction of articles show strong dependence. They argued that fitting a single Markov order to navigation is ill-posed~\cite{arora2022wikipedia}. We do not address order selection. Our history-free model has only first-order parameters, and we test it against held-out humans on group-level properties of trajectories rather than on the likelihood of individual clicks. Like Arora et al., we treat memory as a property of the collective rather than of individual sequences, and like them we find that a first-order model captures a substantial part of human navigation, but not all of it.

The article types in Results Sections~(4)-(5) are one way to understand that which degree leaves unexplained. The types depend on the data, and describe how humans tend to use articles, rather than fixed properties of those articles themselves. Two independent halves of the human data assign the same type to about half of the articles typed in both (Cohen's $\kappa=0.36$), which is more than would be expected if type assignment were random, but it is not perfectly stable. The graph-structure walker, whose residuals are as reproducible as those of humans, assigns types that agree with human ones only at chance ($\kappa=0.00$). 

\subsection*{Article types and the zoom-out conjecture}

West et al. proposed that players first zoom out to a well-connected article and then home in on the target~\cite{west2009wikispeedia,west2012human}. Results Section~(1) seems at first glance to contradict this, since human traversal counts are less correlated with graph measures of connectedness than are any random walker's traversal counts. But these two are not mutually exclusive. A random walk is drawn to well-connected articles (its visit frequencies approximate PageRank), so it is possible that humans pass through well-connected articles early in each game and still are less concentrated there overall, especially if they spend the rest of the game at more specific articles near the target. Because our article types are defined relative to degree, they let us ask which articles might serve as zoom-out points more or less than their connectivity would predict.

Beacons and funnels both draw players from more diverse origins than their in-degree predicts, consistent with the first half of the conjecture, though they differ in what happens next. Trajectories that pass through a beacon connect semantically more distant start and target articles than trajectories that do not, whereas funnels show an effect that is negligible in size (Results Section~(4)). Forks show the same effect as beacons, which suggests that this effect tracks dispersal rather than beacon status. Articles that scatter departing players more than their out-degree predicts are used when the conceptual distance to cover is large, which is what a zoom-out point is for. Corridors show the opposite. They are entered and left along few routes and are used disproportionately when start and target are semantically close, consistent with homing in rather than zooming out. These effects are small and correlational, but they survive controlling for path length. Testing whether funnels channel players toward specific targets would require stratifying Flow Trees by target, which the sparsity of the data does not permit.

Article types cut across topical categories. Countries appear as beacons (``United States''), funnels (``Uganda''), and forks (``Ecuador''); centuries appear as funnels (``19th century'') and corridors (``17th century''). The axis corresponding to the residual of reverse Flow Trees and in-degree is most interpretable, though part of it is driven by graph structure (Results Section (4)). Articles that receive players from more diverse routes than their in-degree predicts tend to be general topics (e.g., ``History of Earth'' or ``Philosophy''), while those reached by fewer routes tend to be specific (e.g., ``Manchester'' or ``Henry VIII of England''). This is consistent with general concepts being reachable from many different conceptual directions.

\subsection*{Connection to landmark navigation and information foraging}
Beacons have a loose parallel in spatial cognition, where certain locations serve as orienting anchors that travelers route through even when doing so lengthens the path~\cite{manley2015shortest}; airlines often route passengers through hub airports, rather than directly to their destination~\cite{guimera2005worldwide}. Trajectories using our beacons are similarly longer (though so are trajectories through any of the four article types), and they are recruited disproportionately when the start and target are far in semantic space.
The parallel is loose, because Wikispeedia has no spatial geometry, and players can only see the text and links on the current article. Planning on a path several clicks ahead therefore requires knowing (or guessing) what lies beyond the current article, and most players have little explicit experience of the game itself. Just under half of the IP hashes played only one game. This suggests that the shared multi-click structure seen in Flow Trees largely comes from what humans intuit about how concepts relate, rather than explicit knowledge of Wikispeedia's underlying network structure. The most frequently used 10\% of IP hashes did contribute to half of all trajectories, so Wikispeedia-specific learning may also be present.

Information foraging theory offers a second perspective~\cite{pirolli1999information}. Players follow the scent of information, moving toward articles that seem to lead toward their target. Our targeted random walker is a simplified implementation of this idea, weighting each step by embedding similarity to the target, but these similarities vary relatively little among the neighbors, so it produces Flow Trees nearly identical to those of the unbiased random walker. This simple form of local semantic gradient-following does not reproduce human navigation. The history-free human model, which follows where players actually click, recovers part of human Flow Tree shape. The scent that players follow is partially present in their first-order click choices, but not in title similarity. This could be through the article text, the placement of links, knowledge that a title's words do not explicitly contain, or structure further ahead than an article's immediate neighbors.

\subsection*{Reader styles}

Our approach complements a parallel line of work, which characterizes Wikipedia navigation by the shape of the knowledge networks readers build. Lydon-Staley et al.~\cite{lydon2021hunters} identified two styles among 149 participants browsing freely: the targeted hunter, who moves through a small set of tightly related articles, and the nomadic busybody, who jumps between loosely connected topics. Zhou et al.~\cite{zhou2024architectural} scaled this to 482,760 mobile readers across 14 languages and 50 countries and found a third, the dancer, who links distant topics through longer creative leaps. Their unit of analysis is the reader and ours is the article, but the underlying strategy is the same, to describe navigation by the topology of what it traverses rather than by which pages get visited. Wikispeedia, by giving players a target, constrains humans to the hunting regime, so the variation we measure is in how articles are used, rather than in how readers peruse the interwebs. 
Targeted and free navigation also differ in shape. Free browsing reconstructed from aggregate click data tends to start from broad articles and narrow steadily, while Wikispeedia players first move from the specific start article to more general ones before narrowing toward the target~\cite{rodi2017search}. Large-scale server logs similarly show that naturally occurring navigation differs from targeted navigation in lab-based settings~\cite{piccardi2023large}.
Whether article types persist across reading styles would require unconstrained browsing data with sufficient traffic per article.
Public clickstream data would not suffice, since trajectories synthesized from it are first-order by construction~\cite{arora2022wikipedia}; the comparison would require sequence-level logs~\cite{piccardi2023large}.

\subsection*{Limitations}
We aggregate across all players, which means the types describe the collective and say nothing about whether individuals differ in how they use these articles. It is possible, for instance, that beacon use is concentrated in a subset of experienced players, and our data cannot distinguish that from uniform adoption. The trajectories we use come from 20,867 unique IP hashes, an imperfect proxy for individuals, just under half of which only played one game (median $2$, mean $3.4$). The most active 10\% of IP hashes account for 49\% of trajectories, and the single most active IP hash played 2,464 games. The aggregate reflects a large population, but we cannot rule out the possibility that enthusiastic power players could have disproportionately shaped it.

Flow Tree construction requires choosing a trajectory length and a sample size. We settled on five decisions and fifteen trajectories after comparing lengths three through six, but this restricts analysis to 815 of the 1000 most-visited articles and necessarily discards games shorter than the window. Results hold on the more skewed articles ranked 1-500 and on the more balanced 501-1000 (see~\nameref{sec:robustness}), which we take as evidence of robustness, but the hyperparameters are not free of consequence.

Because forward and reverse Flow Trees are built from separate sampled sets of trajectories, they do not measure whether the route by which a navigator enters an article predicts the route by which they leave it. The mutual information between incoming and outgoing links~\cite{arora2022wikipedia} measures that dependence directly, though only one click on either side. Relating the two at the article level is a direction of future work.

Our null models vary how the next link is chosen but always choose from the full set of outgoing links. Real players see links laid out on a rendered page, and position, formatting, and prominence within the article all plausibly influence clicks~\cite{dimitrov2017makes,lamprecht2017structure}. The history-free human model used in Results Section (6) partially addresses this concern, since click-preferences are affected by these latent influences and we therefore implicitly absorb them into the click transition matrix without having to specify all possible effects. 

Our language-based random walker null models are simple by design. They rule out these particular language statistics as explanations, but do not rule out semantics in general. The targeted random walker samples in proportion to title similarity rather than choosing the most similar link, which makes it nearly uniform ($H_T=0.996$); agents that greedily follow embedding similarity navigate far more effectively~\cite{ebrahimi2025textual}. The best LLM agents approach human success rates but take more steps than humans~\cite{hickey2026can}. Using such agents as null models in the Flow Tree framework would be a stronger test.

\subsection*{Future directions}

The most direct extension of this work is to examine individual differences in trajectory choices. With per-player trajectory histories it would be possible to ask whether article types are consistent across players or whether the aggregate types average over distinct navigational strategies. Controlled experiments show that such differences exist. In an online Wikipedia navigation game played by 445 participants, performance declined with age and improved with multilingualism~\cite{zhu2024individual}. Pairing a design of this kind with Flow Trees would allow article types to be compared across groups of players. This is not possible with this dataset, because players are identified only by their hashed IP address, an imperfect proxy for individuals, and most contributed only one game.

A second is methodological. We compare Flow Trees through five summary features, but Flow Trees are trees, and tree space has a well-developed geometry. Flow Tree space is Billera–Holmes–Vogtmann (BHV) space quotiented by leaf permutation~\cite{billera2001geometry}, which supplies a metric (see~\nameref{sec:ft-treespace}). Comparing tree shapes directly rather than through feature vectors would make the human-random-walker comparison sharper and would allow clustering articles by tree shape rather than by residual coordinates.

Finally, the forward-reverse $R^{2}$ gap (how much better in-degree predicts reverse Flow Tree features than out-degree predicts those of forward ones; Results Section~(3), Fig~\ref{fig:regression}B) is defined for any directed network with trajectory data, from human mobility traces to animal movement. Whether it tracks other measures of non-reversibility across networks, and whether it identifies the same kind of structural asymmetry we see here, is an open question.

\subsection*{Conclusion}
Collective knowledge is usually inferred from that which a group produces. Flow Trees let us observe it in what a group does, step by step. Null models let us ask how much of it the environment alone would produce. In Wikispeedia, part of what players share is agreement about the next click, and a measurable part extends beyond it.

\section*{Supporting information}
\paragraph*{S1 Appendix.}
\label{sec:ft-feats}
\textbf{Flow Tree features.}
These features are defined and characterized in the context of path Flow Trees in~\cite{bertics2026flowtree}.
Here we define the features we consider below for article Flow Trees, as well as provide a behavioral interpretation in the context of the Wikispeedia dataset (for a graphical depiction of feature definitions, see Fig~\ref{fig:howtobuild}D). For clarity, all interpretation below is written in the context of \textit{forward} Flow Trees, which quantify how a group of trajectories diverges from a shared article. Interpretation in the context of \textit{reverse} Flow Trees applies, substituting ``converge to'' for ``diverge from'', ``entering'' for ``leaving'', and ``end'' for ``start''.
\begin{itemize}
    \item \textbf{Number of Leaves:} 
    \begin{itemize}
        \item \underline{Definition:} Number of distinct trajectories.
        \item \underline{Interpretation:} Diversity of navigation paths leaving an article. Fewer leaves mean more navigators share an identical trajectory leaving the article; many leaves suggest the article acts as a branching point, scattering trajectories. The upper bound is the number of trajectories in the Flow Tree.
    \end{itemize}
    
    \item \textbf{Number of Nodes:} 
    \begin{itemize}
        \item \underline{Definition:} Total number of group split events plus number of groups at the end (i.e., nodes = splits + leaves).
        \item \underline{Interpretation:} How many times the group collectively diverged. Higher values imply greater fragmentation and less shared decision-making across the path.
    \end{itemize}
    
    \item \textbf{Depth:} 
    \begin{itemize}
        \item \underline{Definition:} Length of the longest path from root (shared starting article where all trajectories begin) to leaf on the Flow Tree.
        \item \underline{Interpretation:} The maximum number of split-causing decision points along any single trajectory. Depth is bounded above by the longest trajectory (for analysis in this paper, we truncate all trajectories to 5). Smaller depth can indicate either that most disagreement and divergence occurs immediately, or that there is little disagreement. More depth indicates there are many possible split-causing decision points on one trajectory.
    \end{itemize}
    \item \textbf{Average branching factor:} 
    \begin{itemize}
        \item \underline{Definition:} Average number of child nodes per internal Flow Tree node. 
        \item \underline{Interpretation:} Local decision diversity; average number of distinct choices being made at each split. This is constrained, but not determined, by the topology of the underlying environment. In Wikispeedia, where the out-degree of articles is larger than the number of trajectories we construct Flow Trees from, the upper bound is the number of trajectories in the Flow Tree, in the degenerate case of a fan Flow Tree, where the root node has maximal degree.
    \end{itemize}
     
    \item \textbf{Diameter:} 
    \begin{itemize}
        \item \underline{Definition:} The longest shortest path between two leaf nodes.
        \item \underline{Interpretation:} Maximum behavioral spread across the group; how different the most divergent pair of completed trajectories is. A larger diameter can indicate this article is being used for a variety of navigation tasks.
    \end{itemize}
\end{itemize}

\paragraph*{S2 Appendix.}
\label{sec:forward-reverse-dynamics}
\textbf{Forward and reverse traversal are distinct processes.}
Results Section (3) shows that in-degree predicts reverse Flow Tree shape better than out-degree does forward for humans. The opposite is true for 12 of the 16 random-walker-feature combinations. Before we attribute something about such an asymmetry to the navigator or to its strategy, we need to know how much of the forward–reverse difference could be a property of the graph itself. Here, we establish that forward and reverse traversal are formally distinct processes on a directed graph, we quantify how far Wikispeedia is from being the same forward and backward, and we explain how we constructed our random-walker reverse Flow Trees in a way that is most principled for comparison with humans.

Reverse Flow Trees for random walkers could in principle be constructed in two ways, either by running a forward walker on the reversed graph from each article (single-origin), or by collecting and reversing ordinary forward trajectories that pass through it (multi-origin). 
The second method corresponds to how reverse Flow Trees are constructed using human data. 
These two processes are not mirrors of each other on the Wikispeedia graph. And, more generally, reversing a collection of random walks on a graph $G=(V,E)$ is not the same as generating a collection of random walks on the reversed graph, $G^R=(V,E^R)$, even when matched for starting articles and trajectory length. The two reverse random-walker constructions produce measurably different reverse Flow Trees (see~\nameref{fig:asymmetry}C). 

Reversing a directed graph is a well-defined operation, which turns every outgoing edge into an incoming edge and thus makes the
in-degree distribution of $G$ exactly the out-degree distribution of $G^{R}$. But these are global identities, and a walk is a local process. Formally, the time-reversal of a Markov chain with transition matrix $P$ and stationary distribution $\pi$ has transitions $P^{*}(y \to x) = \pi(x) P(x \to y) / \pi(y)$, which is a simple random walk on $G^{R}$ only when $\pi$ is proportional to out-degree. For an undirected graph this holds, and the two processes are the same. But, generally, for a directed graph it does not. Here, $\pi$ is the PageRank-like stationary distribution, and the reversed chain weights each predecessor by its stationary mass rather than uniformly.

Wikispeedia has different in- vs. out-degree distributions (see~\nameref{fig:asymmetry}A). The forward and reverse stationary distributions are only weakly related across the most-visited articles ($r = 0.29$ on the top 1000, see~\nameref{fig:asymmetry}B). Taxonomic and classification pages such as ``Chordate'' (338 in, 8 out) and ``Scientific classification'' (519 in, 25 out) dominate the forward stationary distribution, while long survey articles such as ``History of the world'' (7 in, 147 out) dominate the reverse one. This makes sense given how Wikipedia articles tend to be written. An article about animals will likely mention the phylum they belong to, so ``Chordate'' accumulates links in. But, the article itself is a short definition that does not link to much. A survey article, in contrast, like ``History of the world'' links out to hundreds of other articles, and it is too general for articles to have reason to link back to it. A forward random walker would end up, therefore, at taxonomic pages, which are easy to enter and hard to leave, which is the same reason why a random walker on the reversed graph would end up at survey pages.

Flow Trees constructed from trajectories generated using these two methods differ significantly for the unbiased walker (single- vs. multi-origin: depth $1.83$ vs.\ $2.01$; branching factor $8.47$ vs.\ $6.61$; all $p<0.001$) but only slightly for the graph-structure walker (depth 2.08 vs. 2.05, diameter 3.90 vs. 3.87, both $p<0.01$ given $n=814$; branching factor 6.12 vs. 6.17, $p=0.22$). One possibility is that degree-weighting on the reversed graph $G^{R}$ approximates the stationary-mass reweighting that formally relates the two processes. This is the same walker that comes closest to humans in both directions in Results Section~(2) and that produces the most reproducible reverse residuals in Results Section~(5). We report multi-origin results throughout, as it is more naturalistic, and have verified that single-origin walkers remain significantly different from humans on every Flow Tree feature, with comparable effect sizes.

{\centering \includegraphics[width=0.8\textwidth,page=8]{figures/ft_wikispeedia.pdf}}
\paragraph*{S2 Fig.}
\label{fig:asymmetry}
{\bf Structural asymmetry of forward vs. reverse traversal in Wikispeedia.} 
A: Distribution of log in-degree (blue) and log out-degree (red) for the 815 analyzed articles. Selecting articles by visit frequency selects for high in-degree, so a walker leaving one of these articles faces a very different local decision space than a walker arriving at it.
B: Log PageRank under a forward walk on $G$ vs. on the reversed graph $G^{R}$. If reversible these would coincide. Walking forward, the stationary distribution concentrates on dead-end pages that have many links in but few out, such as taxonomic pages like ``Chordate'' (338 in, 8 out). Walking forward on the reversed graph, the same dead-end logic applies but to long survey articles with many outgoing links, such as ``History of the world'' (7 in, 147 out).
C: Mean reverse Flow Tree features under the two constructions, collecting and reversing ordinary forward walks on $G$ (pink) vs. running walks on $G^{R}$ from each article (purple), unbiased walker
shown. Error bars are standard errors across articles.

\paragraph*{S3 Appendix.}
\label{sec:ft-treespace}
\textbf{Metric structure of Flow Tree space.}
Equivalently, a Flow Tree is an unlabeled metric $n$-tree, whose topology encodes trajectory splits and whose edge weights encode the group sizes at those splits. An $n$-tree is a tree with a distinguished vertex (the root) and $n$ vertices of degree 1 (the leaves). For an n-tree to be a metric n-tree it must have all interior edge weights (i.e., those not touching a leaf) greater than zero.

A well-defined metric space for trees is BHV space, which provides a notion of distance among semi-labeled metric n-trees~\cite{billera2001geometry}. BHV space is CAT(0), meaning that it is non-positively curved. Fréchet means of trees exist and are unique~\cite{sturm2003probability}. Flow Tree space is BHV space, quotiented by permutation of leaf labels. The quotient is still a metric space, giving a principled distance between Flow Trees, although means in it need not be unique.


\paragraph*{S4 Appendix.}
\label{sec:robustness}
\textbf{Results are robust to unbalanced in-degree-out-degree ratio.}

The main text reports results for 815 of the top 1000 most commonly traversed articles. This set of articles has relatively large in-degree (mean in-degree$=96.5$, mean out-degree$=47.9$). We replicate the core results below for two subsets of these 1000 to mitigate a potential bias of high in-to-out-degree ratio: 458 of the top 500 articles (mean in-degree$=132.4$, mean out-degree$=53.4$) and 357 of the 501st-1000th articles (mean in-degree$=50.4$, mean out-degree$=41.0$), where it is much more balanced. 

For all tables in this Appendix, H=human, U=unbiased random walker, G=graph-structure random walker, S=semantic random walker, T=targeted random walker.

\begin{itemize}
    \item \textbf{Top 500 most commonly traversed articles.}
    \begin{itemize}
        \item Forward Flow Tree features (Table 2):~\nameref{tab:forward_ft_stats-top500}
        \item Reverse Flow Tree features (Table 3):~\nameref{tab:reverse_ft_stats-top500}
        \item Degree regression summary (Table 4):~\nameref{tab:regression-top500}
    \end{itemize}
    \item \textbf{501-1000th most commonly traversed articles.}
    \begin{itemize}
        \item Forward Flow Tree features (Table 2):~\nameref{tab:forward_ft_stats-mid500}
        \item Reverse Flow Tree features (Table 3):~\nameref{tab:reverse_ft_stats-mid500}
        \item Degree regression summary (Table 4):~\nameref{tab:regression-mid500}
    \end{itemize}
\end{itemize}

\paragraph*{S1 Table.}
\label{tab:forward_ft_stats-top500}
\textbf{Mean values of forward Flow Tree features for humans and four random-walker models, top 500 most-traversed articles (article-level averages, $n=458$ articles).} 
Welch's $t$-test $p$-values and Cohen's $d$ are reported for each human vs. random-walker comparison. For Cohen's $d$ effect size: small $|d|>0.2$, medium $|d|>0.5$, large $|d|>0.8$. $^{***}$ $p<0.001$; $p$-values are corrected using the Benjamini \& Hochberg procedure.
\begin{table}[h!]
\centering
\small
\begin{tabular}{lccccccccc}
\toprule
 & \multicolumn{5}{c}{Mean} & \multicolumn{4}{c}{$d$} \\
\cmidrule(lr){2-6}\cmidrule(lr){7-10}
Feature & H & U & G & S & T & \multicolumn{1}{c}{vs.\ U} & \multicolumn{1}{c}{vs. G} & \multicolumn{1}{c}{vs. S} & \multicolumn{1}{c}{vs. T}\\
\midrule
Num. nodes & 18.98 & 18.13 & 18.58 & 18.16 & 18.16 & $+0.84^{***}$ & $+0.42^{***}$ & $+0.81^{***}$ & $+0.83^{***}$ \\
Num. leaves & 14.91 & 15.00 & 15.00 & 15.00 & 15.00 & $-0.32^{***}$ & $-0.32^{***}$ & $-0.32^{***}$ & $-0.32^{***}$ \\
Depth & 2.285 & 1.919 & 1.984 & 1.914 & 1.922 & $+1.39^{***}$ & $+1.22^{***}$ & $+1.41^{***}$ & $+1.40^{***}$ \\
Diameter & 4.184 & 3.554 & 3.769 & 3.563 & 3.575 & $+1.34^{***}$ & $+0.99^{***}$ & $+1.31^{***}$ & $+1.32^{***}$ \\
Branching factor & 5.572 & 7.364 & 6.469 & 7.319 & 7.246 & $-1.05^{***}$ & $-0.63^{***}$ & $-1.00^{***}$ & $-1.00^{***}$ \\
\bottomrule
\end{tabular}
\end{table}

\paragraph*{S2 Table.}
\label{tab:reverse_ft_stats-top500}
\textbf{Mean values of reverse Flow Tree features for humans and four random-walker models, top 500 most-traversed articles (article-level averages, $n=458$ articles).} 
Welch's $t$-test $p$-values and Cohen's $d$ are reported for each human vs. random-walker comparison. For Cohen's $d$ effect size: small $|d|>0.2$, medium $|d|>0.5$, large $|d|>0.8$. $^{***}$ $p<0.001$; $p$-values are corrected using the Benjamini \& Hochberg procedure.
\begin{table}[h!]
\centering
\begin{tabular}{lccccccccc}
\toprule
 & \multicolumn{5}{c}{Mean} & \multicolumn{4}{c}{$d$} \\
\cmidrule(lr){2-6}\cmidrule(lr){7-10}
Feature & H & U & G & S & T & \multicolumn{1}{c}{vs.\ U} & \multicolumn{1}{c}{vs. G} & \multicolumn{1}{c}{vs. S} & \multicolumn{1}{c}{vs. T}\\
\midrule
Num. nodes & 19.17 & 18.16 & 18.41 & 18.30 & 18.18 & $+0.98^{***}$ & $+0.77^{***}$ & $+0.86^{***}$ & $+0.97^{***}$ \\
Num. leaves & 14.92 & 15.00 & 15.00 & 15.00 & 15.00 & $-0.34^{***}$ & $-0.35^{***}$ & $-0.35^{***}$ & $-0.35^{***}$ \\
Depth & 2.448 & 1.937 & 1.984 & 1.963 & 1.946 & $+1.45^{***}$ & $+1.41^{***}$ & $+1.45^{***}$ & $+1.45^{***}$ \\
Diameter & 4.373 & 3.598 & 3.729 & 3.662 & 3.618 & $+1.34^{***}$ & $+1.24^{***}$ & $+1.31^{***}$ & $+1.35^{***}$ \\
Branching factor & 5.446 & 7.311 & 6.667 & 6.992 & 7.218 & $-1.00^{***}$ & $-0.79^{***}$ & $-0.92^{***}$ & $-1.00^{***}$ \\
\bottomrule
\end{tabular}
\end{table}

\paragraph*{S3 Table.}
\label{tab:regression-top500}
\textbf{Degree regression summary for forward and reverse Flow Trees, top 500 most-traversed articles ($n=458$).} 
$R^2_{\mathrm{dom}}$: univariate $R^2$ of the dominant-direction degree
predictor (out-degree for forward, in-degree for reverse).
$R^2_{\mathrm{both}}$: multivariate $R^2$ with both degrees included.
Adding the opposite-direction degree contributes negligibly for all walker
types ($\Delta R^2 \leq 0.015$).
Number of leaves is also omitted, since random-walker variance is near zero (all trajectories produce the maximum of 15 distinct paths),
making degree-based prediction uninformative.
\begin{table}[h!]
\centering
\begin{tabular}{lcccccccccc}
\toprule
 & \multicolumn{5}{c}{$R^2_{\mathrm{dom}}$} & \multicolumn{5}{c}{$R^2_{\mathrm{both}}$} \\
\cmidrule(lr){2-6}\cmidrule(lr){7-11}
Feature & H & U & G & S & T & H & U & G & S & T \\
\midrule
\textit{Forward (out-deg.)} \\
\hspace{0.2cm}Num. nodes & 0.307 & 0.799 & 0.736 & 0.812 & 0.803 & 0.314 & 0.799 & 0.736 & 0.813 & 0.803 \\
\hspace{0.2cm}Depth & 0.092 & 0.550 & 0.449 & 0.528 & 0.488 & 0.099 & 0.551 & 0.449 & 0.528 & 0.488 \\
\hspace{0.2cm}Diameter & 0.197 & 0.694 & 0.634 & 0.700 & 0.695 & 0.205 & 0.694 & 0.634 & 0.700 & 0.696 \\
\hspace{0.2cm}Branching factor & 0.307 & 0.718 & 0.658 & 0.722 & 0.698 & 0.322 & 0.718 & 0.658 & 0.724 & 0.699 \\
\midrule
\textit{Reverse (in-deg.)} \\
\hspace{0.2cm}Num. nodes & 0.325 & 0.732 & 0.537 & 0.661 & 0.699 & 0.327 & 0.732 & 0.539 & 0.661 & 0.700 \\
\hspace{0.2cm}Depth & 0.214 & 0.580 & 0.309 & 0.462 & 0.570 & 0.214 & 0.581 & 0.313 & 0.465 & 0.570 \\
\hspace{0.2cm}Diameter & 0.309 & 0.714 & 0.528 & 0.615 & 0.686 & 0.310 & 0.715 & 0.531 & 0.617 & 0.688 \\
\hspace{0.2cm}Branching factor & 0.366 & 0.714 & 0.515 & 0.639 & 0.688 & 0.370 & 0.715 & 0.519 & 0.642 & 0.688 \\
\bottomrule
\end{tabular}
\end{table}

\paragraph*{S4 Table.}
\label{tab:forward_ft_stats-mid500}
\textbf{Mean values of forward Flow Tree features for humans and four random-walker models, 501st-1000th most-traversed articles (article-level averages, $n=357$ articles).} 
Welch's $t$-test $p$-values and Cohen's $d$ are reported for each human vs. random-walker comparison.
For Cohen's $d$ effect size: small $|d|>0.2$, medium $|d|>0.5$, large $|d|>0.8$.
$^{***}$ $p<0.001$; $p$-values are corrected using the Benjamini \& Hochberg procedure.
\begin{table}[h!]
\centering
\small
\begin{tabular}{lccccccccc}
\toprule
 & \multicolumn{5}{c}{Mean} & \multicolumn{4}{c}{$d$} \\
\cmidrule(lr){2-6}\cmidrule(lr){7-10}
Feature & H & U & G & S & T & \multicolumn{1}{c}{vs.\ U} & \multicolumn{1}{c}{vs. G} & \multicolumn{1}{c}{vs. S} & \multicolumn{1}{c}{vs. T}\\
\midrule
Num. nodes & 19.12 & 18.48 & 18.86 & 18.47 & 18.46 & $+0.56^{***}$ & $+0.23^{**}$ & $+0.57^{***}$ & $+0.58^{***}$ \\
Num. leaves & 14.86 & 15.00 & 15.00 & 15.00 & 15.00 & $-0.32^{***}$ & $-0.32^{***}$ & $-0.32^{***}$ & $-0.32^{***}$ \\
Depth & 2.308 & 1.980 & 2.027 & 1.968 & 1.990 & $+1.10^{***}$ & $+0.97^{***}$ & $+1.15^{***}$ & $+1.07^{***}$ \\
Diameter & 4.243 & 3.709 & 3.881 & 3.706 & 3.732 & $+1.15^{***}$ & $+0.86^{***}$ & $+1.17^{***}$ & $+1.11^{***}$ \\
Branching factor & 5.334 & 6.637 & 5.954 & 6.673 & 6.615 & $-0.86^{***}$ & $-0.50^{***}$ & $-0.89^{***}$ & $-0.87^{***}$ \\
\bottomrule
\end{tabular}
\end{table}

\paragraph*{S5 Table.}
\label{tab:reverse_ft_stats-mid500}
\textbf{Mean values of reverse Flow Tree features for humans and four random-walker models, 501st-1000th most-traversed articles (article-level averages, $n=357$ articles).} 
Welch's $t$-test $p$-values and Cohen's $d$ are reported for each human vs. random-walker comparison.
For Cohen's $d$ effect size: small $|d|>0.2$, medium $|d|>0.5$, large $|d|>0.8$.
$^{***}$ $p<0.001$; $p$-values are corrected using the Benjamini \& Hochberg procedure.
\begin{table}[h!]
\centering
\begin{tabular}{lccccccccc}
\toprule
 & \multicolumn{5}{c}{Mean} & \multicolumn{4}{c}{$d$} \\
\cmidrule(lr){2-6}\cmidrule(lr){7-10}
Feature & H & U & G & S & T & \multicolumn{1}{c}{vs.\ U} & \multicolumn{1}{c}{vs. G} & \multicolumn{1}{c}{vs. S} & \multicolumn{1}{c}{vs. T}\\
\midrule
Num. nodes & 19.72 & 19.05 & 19.08 & 19.19 & 19.09 & $+0.67^{***}$ & $+0.65^{***}$ & $+0.54^{***}$ & $+0.62^{***}$ \\
Num. leaves & 14.84 & 15.00 & 15.00 & 15.00 & 15.00 & $-0.40^{***}$ & $-0.40^{***}$ & $-0.40^{***}$ & $-0.40^{***}$ \\
Depth & 2.704 & 2.113 & 2.128 & 2.131 & 2.109 & $+1.71^{***}$ & $+1.63^{***}$ & $+1.62^{***}$ & $+1.73^{***}$ \\
Diameter & 4.754 & 3.997 & 4.040 & 4.046 & 4.000 & $+1.54^{***}$ & $+1.48^{***}$ & $+1.44^{***}$ & $+1.55^{***}$ \\
Branching factor & 4.850 & 5.703 & 5.531 & 5.526 & 5.636 & $-0.85^{***}$ & $-0.87^{***}$ & $-0.76^{***}$ & $-0.83^{***}$ \\
\bottomrule
\end{tabular}
\end{table}

\paragraph*{S6 Table.}
\label{tab:regression-mid500}
\textbf{Degree regression summary for forward and reverse Flow Trees, 501st-1000th most-traversed articles ($n=357$).} 
$R^2_{\mathrm{dom}}$: univariate $R^2$ of the dominant-direction degree
predictor (out-degree for forward, in-degree for reverse).
$R^2_{\mathrm{both}}$: multivariate $R^2$ with both degrees included.
Adding the opposite-direction degree contributes negligibly for all walker
types ($\Delta R^2 \leq 0.011$).
Number of leaves is also omitted, since random-walker variance is near zero (all trajectories produce the maximum of 15 distinct paths),
making degree-based prediction uninformative.
\begin{table}[h!]
\centering
\begin{tabular}{lcccccccccc}
\toprule
 & \multicolumn{5}{c}{$R^2_{\mathrm{dom}}$} & \multicolumn{5}{c}{$R^2_{\mathrm{both}}$} \\
\cmidrule(lr){2-6}\cmidrule(lr){7-11}
Feature & H & U & G & S & T & H & U & G & S & T \\
\midrule
\textit{Forward (out-deg.)} \\
\hspace{0.2cm}Num. nodes & 0.105 & 0.792 & 0.675 & 0.766 & 0.738 & 0.110 & 0.793 & 0.675 & 0.766 & 0.740 \\
\hspace{0.2cm}Depth & 0.015 & 0.420 & 0.317 & 0.433 & 0.364 & 0.019 & 0.423 & 0.320 & 0.437 & 0.365 \\
\hspace{0.2cm}Diameter & 0.066 & 0.646 & 0.518 & 0.618 & 0.573 & 0.068 & 0.648 & 0.522 & 0.623 & 0.574 \\
\hspace{0.2cm}Branching factor & 0.155 & 0.672 & 0.612 & 0.666 & 0.601 & 0.155 & 0.673 & 0.615 & 0.667 & 0.604 \\
\midrule
\textit{Reverse (in-deg.)} \\
\hspace{0.2cm}Num. nodes & 0.121 & 0.519 & 0.296 & 0.550 & 0.562 & 0.124 & 0.523 & 0.296 & 0.558 & 0.570 \\
\hspace{0.2cm}Depth & 0.098 & 0.284 & 0.189 & 0.287 & 0.328 & 0.098 & 0.284 & 0.189 & 0.287 & 0.329 \\
\hspace{0.2cm}Diameter & 0.124 & 0.453 & 0.295 & 0.459 & 0.456 & 0.125 & 0.456 & 0.296 & 0.460 & 0.457 \\
\hspace{0.2cm}Branching factor & 0.149 & 0.468 & 0.315 & 0.510 & 0.476 & 0.156 & 0.479 & 0.316 & 0.519 & 0.486 \\
\bottomrule
\end{tabular}
\end{table}






\FloatBarrier
\section*{Acknowledgments}
We thank Dale Zhou and Aaron Bornstein for useful discussions and feedback. This research was supported by NIH R01NS119468 and the UCSB Chancellor's Fellowship.


%
%
%
\bibliography{ref}






\end{document}